\documentclass[a4paper,fleqn]{cas-dc}

\usepackage[authoryear,longnamesfirst]{natbib}
\usepackage{amsmath,amssymb,amsthm,bm,graphicx,booktabs,array,tabularx}
\usepackage{algorithm,algorithmic}
\usepackage{tikz}
\usepackage{url}
\ExplSyntaxOn
\file_if_exist:nF { thumbnails/cas-email.jpeg }
  { \bool_gset_true:N \g_stm_nologo_bool }
\ExplSyntaxOff

\definecolor{QELCitationBlue}{RGB}{0,55,180}
\definecolor{QELQAColor}{RGB}{170,0,0}
\hypersetup{citecolor=QELCitationBlue,hypertexnames=false}
\newtheorem{theorem}{Theorem}
\newtheorem{proposition}{Proposition}
\newtheorem{lemma}{Lemma}
\newtheorem{corollary}{Corollary}
\theoremstyle{definition}
\newtheorem{definition}{Definition}
\newtheorem{example}{Example}
\theoremstyle{remark}
\newtheorem*{remark}{Remark}
\newtheorem{assumption}{Assumption}   
\newcommand{\boldBox}{\Box}
\newcommand{\mdiamond}{\Diamond}
\newcommand{\semantics}[1]{[\![#1]\!]}
\newcommand{\EL}{\textsf{EL}}

\newcommand{\TES}{\textsf{LINE}}
\newcommand{\IE}{\textsf{IE}}

\newcommand{\R}{\mathbb{R}}
\newcommand{\Q}{\mathbb{Q}}
\newcommand{\Nat}{\mathbb{N}}

\begin{document}
\let\WriteBookmarks\relax
\def\floatpagepagefraction{1}
\def\textpagefraction{.001}

\shorttitle{Rational Ensemble Logic for Cohort Discovery}

\shortauthors{Zhang et al.}

\title[mode=title]{Rational Ensemble Logic for Cohort Discovery with Real-World Electrophysiology Data}

\author[1]{Guo-Qiang Zhang}
\cormark[1]
\ead{Guo-Qiang.Zhang@uth.tmc.edu}
\credit{Conceptualization, Methodology, Formal analysis, Writing-original draft \& editing}

\author[1]{Xubing Hao}
\credit{Investigation, Methodology, Writing-review \& editing}

\author[1]{Yan Huang}
\credit{Investigation, Methodology, Writing-review \& editing}

\author[1]{Xiaojin Li}
\credit{Investigation, Methodology, Writing-review \& editing}

\author[1]{Licong Cui}
\credit{Investigation, Writing-review \& editing}

\affiliation[1]{organization={The University of Texas Health Science Center at Houston},
            addressline={7000 Fannin Street},
            city={Houston},
            postcode={77030},
            state={Texas},
            country={USA}}

\cortext[1]{Corresponding author. ORCID: 0000-0002-3663-1109}

\begin{abstract}
\textbf{\bf Objective:}
Biomedical knowledge representation should make the passage from observations to conclusions transparent, rigorous, understandable, explainable, and reproducible (TRUER).
We introduce Rational Ensemble Logic (QEL) for specifying temporal phenotypes in real-world electrophysiology data (RWED), with a meaning independent of the implementation software.
\textbf{Materials and methods:} QEL combines three parameterized modal operators (exact displacement, bounded existence, and bounded universality)
 over rational time with Boolean connectives and first-order quantification.
Its Labeled Interval Ensemble (LINE) semantics represents annotated records as finite interval ensembles.
We relate the operators to morphological operations and develop endpoint-array algorithms and finite rational evidence sets for model checking.
\textbf{Results:} We prove closure and semantic correctness of the interval operations, obtain finite evidence sets by quantifier elimination, and establish polynomial data complexity for fixed formulas.
On 54 manually constructed records, a prototype agreed with the assigned outcomes for base and strict postictal generalized EEG suppression (PGES) queries.
A separate feasibility experiment over 1,976 patients identified 37 qualifying patients and 45 annotation patterns in 16 seconds.
The released prototype uses endpoint candidates and numerical tolerances. 
\textbf{Conclusions:} QEL supports the TRUER principle through explicit phenotype definitions, precise semantics, and provably correct evaluation procedures, with rational witnesses for existential queries.
These provide a logical foundation for formal specification and correctness of algorithms that are independent of specific clinical interpretation, data annotation quality, and software implementation.

\end{abstract}


\maketitle
\section{Introduction}
\label{intro}

Continuously recorded signals such as the electroencephalogram (EEG), electrocardiogram
(ECG), and electromyogram (EMG) occur in resources including
PhysioBank~\citep{moody2001physionet,moody2022physionet}, the National Sleep Research Resource
(NSRR)~\citep{nsrr}, and UK Biobank~\citep{bycroft2018uk}.
Annotations label segments with clinical events and states. These annotations support phenotype
definitions in terms of the occurrence, frequency, duration, ordering, and cadence of events.
Identifying postictal generalized EEG suppression (PGES) after a generalized tonic--clonic (GTC)
seizure, for example, requires specifying the relation between seizure termination and suppression
onset and duration. We take the annotations as given, whether produced by experts or automated
detectors; their quality and conventions may differ across datasets.

Cohort discovery identifies records exhibiting a specified phenotype. Database queries and
application-specific scripts can implement temporal criteria, but the assumptions and logical
relationships may remain implicit in the code. A representation that is both human-readable and
computable is a stated desideratum for phenotype algorithms~\citep{mo2015desiderata}.
Existing biomedical formalisms serve distinct purposes: SNOMED CT and
UMLS~\citep{bodenreider2004umls} organize clinical terminology;
SBML~\citep{hucka2003sbml} and BioPAX~\citep{demir2010biopax} represent biochemical models and
pathways; temporal logics~\citep{pnueli1977temporal} specify behavior over time; and mathematical
morphology describes transformations of spatial
sets~\citep{serra1982morphology,matheron1975random,heijmans1994morphological}.
Our concern is a common logical language in which metric relationships and quantifier scope are
explicit and support compositional reasoning.

Biomedical science has entered an era of
data abundance, yet the knowledge derived from those data is not inherently 
TRUER: Transparent, Rigorous, Understandable, Explainable, and Reproducible
(Table~\ref{tab:truer}). 
Meaning remains fragmented across modality-specific
representations, analytic pipelines, software systems, and human interpretation. In current
practice, a phenotype is the output of a pipeline: a query bound to one institution’s data model, a number
produced by segmentation code, an annotation that depends on the scorer. Large language models (AI) produce
fluent conclusions whose reasoning and provenance remain opaque and raise reproducibility and
the accessibility concerns~\citep{haibekains2020,mcdermott2021}.
 Ontologies supply semantic
precision without the machinery for deduction over heterogeneous data; knowledge graphs connect
entities, while rigorous reasoning over them at scale remains open.

\citet{Gille2020Trust} distinguish trust in AI systems from the properties that make them
trustworthy and argue that technical explainability alone cannot establish trust in healthcare AI.
Knowledge representations should therefore expose their assumptions, semantics, and reasoning
for examination~\citep{rudin2019stop}. TRUER treats these properties as requirements of the
representation and reasoning framework, as well as of its evaluation. Whereas the FAIR
principles~\citep{wilkinson2016fair} concern data stewardship, TRUER concerns the knowledge
derived from data.

\begin{table*}[t]
\caption{The TRUER principle: the meaning of each property, how Rational Ensemble Logic (QEL)
supports it, and where it is addressed.}\label{tab:truer}
\centering
\footnotesize
\renewcommand{\arraystretch}{1.25}
\setlength{\tabcolsep}{4.5pt}
\begin{tabularx}{\textwidth}{|>{\bfseries}l|>{\raggedright\arraybackslash}p{0.235\textwidth}|>{\raggedright\arraybackslash}X|>{\raggedright\arraybackslash}p{0.215\textwidth}|}
\hline
Property & \textbf{Meaning} & \textbf{How QEL supports it} & \textbf{Where shown} \\ \hline
Transparent & The assumptions and the logical chain of reasoning for a biomedical conclusion are explicitly inspectable. & \emph{By design.} A phenotype definition is a single declarative formula. Each temporal assumption (ordering, duration, latency, persistence) appears as an operator, a bound, or a quantifier; the semantics determines their interpretation. & GTC and PGES specifications (Sections~\ref{gtc-qel} and~\ref{subsec:pges}). \\ \hline
Rigorous & Biomedical representations and conclusions are governed by precise definitions, validated methods, explicit assumptions, and verifiable evidence. & \emph{By construction.} Every formula has a mathematically defined meaning over annotated records, and the evaluation algorithms are proved to compute exactly that meaning, with stated complexity. & Semantics (Definitions~\ref{qelsemantics}, \ref{tes-semantics}); correctness (Propositions~\ref{prop:ie-closure}, \ref{closed}; Theorems~\ref{thm:finite-test-set}, \ref{complexity}); complexity (Theorem~\ref{complexity}). \\ \hline
Understandable & The meaning of a representation or result is accessible to clinicians and investigators who are in a position to interpret, evaluate, and use it. & \emph{By design.} The three operators read the way temporal criteria are phrased in clinical practice: at this offset, somewhere within this window, throughout this window. In the Cohort Builder, a specification is composed along a timeline, and its formula is synthesized and displayed. & Operators by example (Section~\ref{subsec:operators}) and formally (Definition~\ref{qelsemantics}); Cohort Builder (Section~\ref{subsec:builder}). \\ \hline
Explainable & While understandability clarifies meaning, explainability identifies the reasoning steps from which the conclusion is deduced (transparency is a prerequisite, not a substitute). & \emph{By construction.} An existential specification admits witness values that can be checked against the formula and annotations. The prototype displays matched patterns; its search limitations are stated separately. & Evidence sets (Theorems~\ref{thm:finite-test-set} and~\ref{complexity}); matched annotation patterns returned (Section~\ref{subsec:cohort-generation}) and displayed (Section~\ref{subsec:builder}). \\ \hline
Reproducible & An independent party or context can reconstruct, re-execute, test, and modify a process under stated conditions. & \emph{By design and by construction.} \emph{Reconstruct:} the formula and annotation conventions determine a meaning independently of software. \emph{Re-execute:} conforming evaluators return the same cohort from the same records. \emph{Test:} the prototype is compared with assigned outcomes on constructed records. \emph{Modify:} formula and boundary-convention changes are explicit in the base and strict PGES queries. & PGES specification (Section~\ref{subsec:pges}); strict variant, conformance assessment on 54 records, and released checker (Section~\ref{sec:synthetic-evaluation}); feasibility experiment over 1,976 patients (Section~\ref{subsec:builder}). \\ \hline
\end{tabularx}
\end{table*}

Reproducibility includes more than re-executing shared code on shared data
(Table~\ref{tab:truer}, last row). Two implementations of an informal criterion may disagree
silently. Explicit semantics instead determines what each implementation must compute. Given the
same records, annotation conventions, and reference points, conforming evaluators must return the
same truth values. The definition supplies a common standard against which implementations can
be checked; agreement on selected examples alone does not establish conformance.

Ensemble Logic (EL)~\citep{zhang2024temporal} was motivated by the need for such a language.
Its three primitive operators, exact displacement ($\varphi_u$), bounded existence
($\mdiamond_t\varphi$), and bounded universality ($\boldBox_t\varphi$), express ``at this
offset,'' ``somewhere within this window,'' and ``throughout this window.'' Boolean connectives
and first-order quantification combine these operators into specifications over metric domains
such as time, space, and genomic coordinates. EL is a form of symbolic AI that can reason over
annotations supplied by machine-learning models. The broader question is whether it can support
\emph{Symbolic Biomedicine}: biomedical knowledge represented in a common formal language that
supports the TRUER principle.

This paper studies temporal phenotypes in RWED. Its standpoint is observational rather than
mechanistic~\citep{winskel,pepa,brane,regev}: it describes annotated event
patterns~\citep{liu2022real,lavertu2021new}, without modeling their biological, environmental, or
behavioral causes. We develop Rational Ensemble Logic (QEL), the rational-time instance of EL.
Rational numbers represent stored sampling intervals and annotation timestamps exactly. QEL
therefore permits dense-time reasoning without imposing a fixed sampling grid, while retaining
the temporal relationships recorded in the annotations.

QEL serves as a description language for temporal phenotypes; model checking relates those
descriptions to records. We use \emph{model checking} in its logical and database-theoretic
sense: deciding whether a given structure satisfies a given
formula~\citep{vardi1982complexity,halpern1991model,libkin2004elements,flum2006parameterized}.
The task differs from verifying every behavior of a system
design~\citep{clarke1999model,baier2008principles}. Each annotated record is represented by a
Labeled Interval Ensemble (LINE), and a QEL formula specifies the phenotype. The cohort consists
of records satisfying that formula. Table~\ref{tab:truer} relates this formal development and
the prototype assessment to the five TRUER properties.

\subsection{Contributions}
\label{subsec:contributions}

The paper proposes TRUER as a guiding principle and develops the QEL semantics and algorithms
needed to support it. The formal contribution combines LINE representations, morphological
operations, and quantifier elimination to obtain exact dense-time evaluation with polynomial
data complexity for fixed formulas. Relative to the general
EL framework~\citep{zhang2024temporal} and its discrete
instance~\citep{droste2026discrete}, the new results concern LINE semantics and its computational
consequences (C2--C4).

\begin{description}
\item[C1 (Principle).] The TRUER principle and its realization for temporal phenotypes
  (Table~\ref{tab:truer}), placing QEL as a formal framework for biomedical knowledge representation that makes the passage from observations to conclusions using explicit, mathematically defined specifications.
\item[C2 (Semantics).] A LINE semantics for QEL over rational time, representing annotated
  records as finite interval ensembles (Section~\ref{sec:qel-rational}).
\item[C3 (Morphological correspondence).] The correspondence between QEL operators and
  the geometric operators of displacement, dilation, erosion, with closure and semantic-correctness results
  (Section~\ref{sec:qel-rational}).
\item[C4 (Computation).] Finite rational evidence sets for quantified specifications by quantifier
  elimination (Theorem~\ref{thm:finite-test-set}); endpoint-array algorithms with polynomial data
  complexity (Section~\ref{mc}, Theorem~\ref{complexity}); and, separately, a closure result for
  Euclidean spatial supports (Theorem~\ref{thm:sa-closure}).
\item[C5 (Application prototype).] GTC seizure and PGES specifications; a controlled
  assessment of a prototype on 54 manually constructed LINE records, comparing its outputs with
  assigned outcomes for specified temporal patterns; a cohort-discovery
  feasibility experiment over 1,976 patients; and an interactive Cohort Builder
  (Section~\ref{sec:application}).
\end{description}

\subsection{Organization}
\label{subsec:organization}

Section~\ref{sec:background} introduces the clinical example and related formalisms.
Section~\ref{metricmon} defines EL and its rational-time instance QEL.
Section~\ref{sec:qel-rational} develops LINE semantics, morphological operations, and finite
evidence sets. Section~\ref{mc} gives the algorithms and their complexity.
Section~\ref{sec:application} presents the GTC and PGES specifications, the prototype assessment,
and the Cohort Builder. Section~\ref{sec:conclusion} discusses the results and limitations.
The Appendix proves Theorems~\ref{thm:sa-closure}, \ref{thm:finite-test-set}, and~\ref{complexity}
and establishes the combined-complexity and satisfiability results.
\section{Background}\label{sec:background}

We introduce annotated records and cohort discovery through a seizure example, illustrate the
three EL operators, and compare QEL with temporal query systems and related logics.

\subsection{Annotated Records and Cohort Discovery}\label{subsec:records}

Real-world data are collected in routine care, and regulators and investigators increasingly rely
on the real-world evidence drawn from them \citep{concato2022real,us2018framework}. RWED are
real-world data with two defining attributes: they are temporal, and they are annotated. An
annotation labels an interval of a recording with a clinical event or state, and a \emph{record}
is the collection of annotations of one patient or recording session.

\begin{figure}[pos=tp]
\centering
\begin{tikzpicture}[x=0.0071428571\linewidth,y=1cm,font=\footnotesize]
\path[use as bounding box] (-7,-0.75) rectangle (133,3.70);
\tikzset{phase/.style={font=\sffamily\bfseries\fontsize{5.5}{6.5}\selectfont,text=white}}
\fill[black!32] (-5,3.30) rectangle (10,3.80);  \node[phase,align=center] at (2.5,3.55)  {Preictal\\[-1pt]\scalebox{0.72}{(Hours - Days)}};
\fill[black!55] (10,3.30)  rectangle (30,3.80);  \node[phase,align=center] at (20,3.55)  {Tonic Phase\\[-1pt]\scalebox{0.72}{($\sim$10 - 30 sec)}} ;
\fill[black!77] (30,3.30)  rectangle (80,3.80);  \node[phase,align=center] at (55,3.55)  {Clonic Phase\\[-1pt]\scalebox{0.72}{($\sim$30 - 90 sec)}};
\fill[black!32] (80,3.30)  rectangle (120,3.80); \node[phase,align=center] at (100,3.55) {Postictal Suppression\\[-1pt]\scalebox{0.72}{($\sim$10 - 120 sec)}};
\IfFileExists{GTC_v5.png}%
  {\node[anchor=south west,inner sep=0] at (-5.5,2.30)
     {\includegraphics[width=0.9192\linewidth,height=0.95cm]{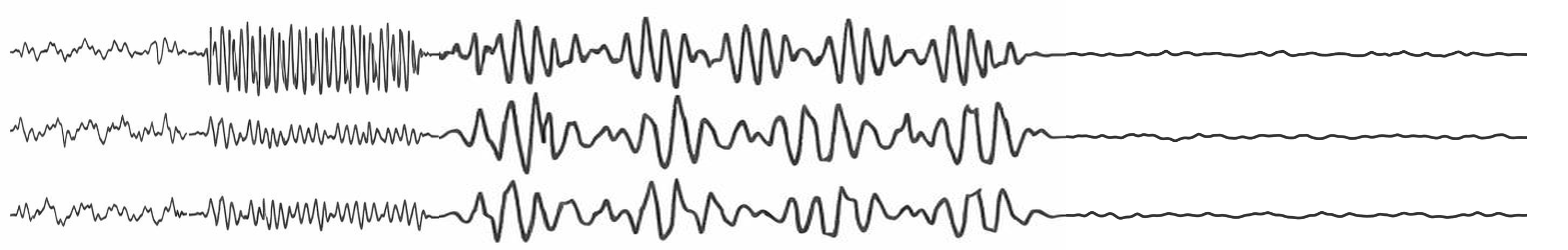}};}%
  {\draw[black!30] (-5,2.30) rectangle (120,3.25);
   \node[font=\tiny] at (57.5,2.775) {Place \texttt{GTC\_waveform\_v5.png} in the compilation directory.};}
\foreach \t in {-5,10,30,80,120} {\draw[black!25,densely dashed] (\t,3.30) -- (\t,2.22);}
\draw[black!55,->] (-7,2.22) -- (130,2.22);
\node[anchor=south,inner sep=1pt,font=\sffamily\itshape\tiny] at (127,2.24) {time};
\draw[->,thick] (10,1.98) -- (10,2.18);
\node[font=\sffamily\itshape\tiny] at (10,1.88) {seizure onset};
\draw[->,thick] (80,1.98) -- (80,2.18);
\node[font=\sffamily\itshape\tiny] at (80,1.88) {seizure end};
\fill[black!20] (10,1.10) rectangle (80,1.44);   \node at (45,1.27) {$S$};
\fill[black!8]  (80,1.10) rectangle (120,1.44);  \draw[black!45] (80,1.10) rectangle (120,1.44);
\node at (100,1.27) {$P$};
\fill[black!42] (10,0.66) rectangle (30,1.00);   \node at (20,0.83) {$T$};
\fill[black!30] (30,0.66) rectangle (80,1.00);   \node at (55,0.83) {$C$};
\draw[->] (0,0.45) -- (130,0.45);
\foreach \t in {0,10,30,80,120} {\draw (\t,0.45) -- (\t,0.39) node[below,inner sep=1.5pt] {\t};}
\draw[|-|] (0,-0.30) -- (10,-0.30)  node[midway,below,inner sep=3pt] {$x$};
\draw[|-|] (10,-0.30) -- (30,-0.30) node[midway,below,inner sep=3pt] {$a$};
\draw[|-|] (30,-0.30) -- (80,-0.30) node[midway,below,inner sep=3pt] {$b$};
\draw[|-|,densely dashed] (80,-0.30) -- (120,-0.30) node[midway,below,inner sep=3pt] {$c$};
\end{tikzpicture}
\caption{A GTC seizure as a signal and as a record. Top: waveform rendering. Bottom: an
illustrative record (times in seconds) with annotations for the seizure ($S$), its tonic ($T$)
and clonic ($C$) phases, and postictal suppression ($P$). The spans $x$, $a$, $b$ form a witness
for $\varphi$ (Section~\ref{subsec:operators}); $c$ extends it to PGES.}
\label{fig:gtc}
\vspace{-1\baselineskip}
\end{figure}
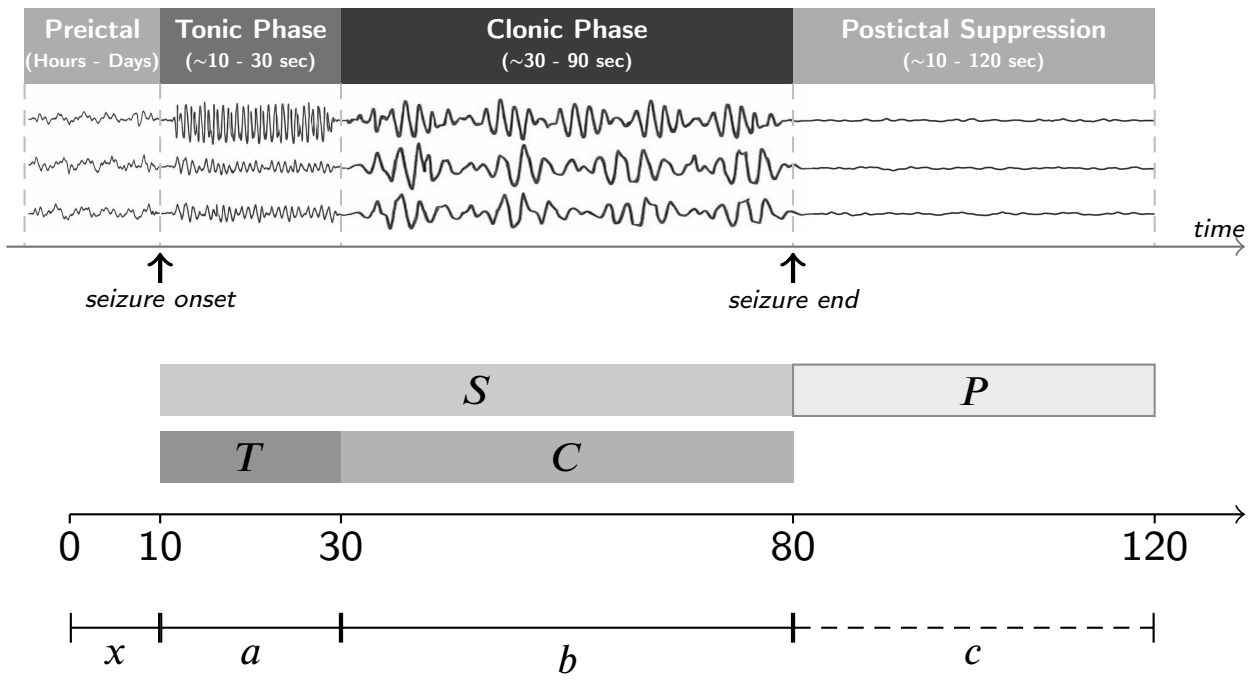

Our running example is the GTC seizure, a major risk marker for sudden unexpected death in
epilepsy \citep[SUDEP;][]{lancet}. Clinically, a GTC seizure has a tonic phase followed by clonic
muscle contractions, and it typically culminates in a prolonged period of unresponsiveness and
confusion~\citep{kodankandath}. The EEG (Figure~\ref{fig:gtc}, top) passes through correspondingly
stereotyped phases, with generalized fast activity during tonic stiffening and spike-wave rhythms
during clonic jerking~\citep{harrison}. Finding the recordings in which a GTC seizure is followed
by PGES is the cohort-discovery problem studied in Section~\ref{sec:application}.

The annotations determine the record. Figure~\ref{fig:gtc} (bottom) shows an
illustrative one, with times in seconds: the seizure ($S$) is annotated on $[10,80)$, its tonic
phase ($T$) on $[10,30)$, its clonic phase ($C$) on $[30,80)$, and postictal suppression ($P$) on
$[80,120)$. Formally, a record assigns to each label a finite set of intervals; this is the LINE
of Section~\ref{sec:qel-rational}.

Cohort discovery can now be stated. A formula defines the phenotype, and model checking determines
which annotated records satisfy it. For existential specifications, a positive answer can be
accompanied by witness values and the corresponding annotations. Sections~\ref{metricmon}--\ref{mc} make this precise and computable.

\subsection{The Three Operators by Example}
\label{subsec:operators}

A formula is evaluated at a time point of a record. A label such as $S$ is true at the time points
that an $S$ annotation covers. Clinical criteria state that an event occurs \emph{at} a given
time, \emph{within} a time window, or \emph{throughout} it, and EL has one operator for
each~\citep{zhang2024temporal}:
\begin{description}\setlength{\itemsep}{0pt}\setlength{\parskip}{0pt}
\item[$\varphi_u$] (exact displacement): $\varphi$ is true $u$ time units later;
\item[$\mdiamond_t\varphi$] (bounded existence): $\varphi$ is true at some time point within the
  next $t$ units;
\item[$\boldBox_t\varphi$] (bounded universality): $\varphi$ is true at every time point within
  the next $t$ units.
\end{description}
At a time point $r$, ``within the next $t$ units'' means the window $[r,r+t)$.

On the record of Figure~\ref{fig:gtc}, take time 10 as the current time point. There
$\boldBox_{70}S$ is true, because $S$ covers $[10,80)$; $\mdiamond_{20}T$ is true, because $T$
occurs within $[10,30)$; and $(\mdiamond_{50}C)_{20}$ is true, because 20\,s later, at time 30,
$C$ occurs within $[30,80)$. Hence the conjunction
\[
\boldBox_{70}S\wedge\mdiamond_{20}T\wedge(\mdiamond_{50}C)_{20}
\]
is true at time 10: a seizure that persists for 70\,s, with tonic activity in its first 20\,s and
clonic activity in 50\,s after.

These constants describe one possible seizure pattern. To allow variable onset and duration,
QEL replaces them with quantified variables:
\[
\varphi:=\exists x\,\exists a\,\exists b\;[\,\boldBox_{a+b}S\wedge\mdiamond_aT\wedge(\mdiamond_bC)_a\,]_x .
\]
At reference time 0, $\varphi$ states that some rational offset $x$ and positive rational
durations $a,b$ determine a seizure window of length $a+b$, with tonic activity somewhere in
its first $a$ units and clonic activity somewhere in its final $b$ units. The record of
Figure~\ref{fig:gtc} satisfies $\varphi$: the values $x=10$, $a=20$, $b=50$ form a \emph{witness}.
Because $a$ and $b$ occur in several operators, the three windows fit together: the clonic window
begins where the tonic window ends, and the seizure spans both. Adding the conjunct
$(\boldBox_cP)_{a+b}$, with $c>0$ quantified, requires suppression to persist for $c$ from the end
of that span; this is the PGES specification of Section~\ref{sec:application}.

Quantifiers also separate criteria that natural language blurs. With time in days, ``fever ($F$)
within two days after every vaccination ($V$)'' is $\forall x\,(V_x\rightarrow(\mdiamond_2F)_x)$,
whereas ``fever within two days after some vaccination'' is
$\exists x\,(V_x\wedge(\mdiamond_2F)_x)$. Here $V$ marks vaccination time points; the window includes that point and excludes its
two-day upper boundary. The formulas define different cohorts.

The operators are specific neither to seizures nor to time. They have been used to formalize
clinical-trial protocols and sleep-scoring rules (Section~\ref{sec:broader-applications}). 

\begin{table*}[t]
\caption{QEL and the temporal formalisms closest to it.}
\label{tab:logic-comparison}
\centering
\footnotesize
\begin{tabularx}{\textwidth}{@{}>{\raggedright\arraybackslash}p{0.15\textwidth}>{\raggedright\arraybackslash}p{0.21\textwidth}>{\raggedright\arraybackslash}p{0.28\textwidth}>{\raggedright\arraybackslash}X@{}}
\toprule
\textbf{Formalism} & \textbf{Evaluated at} & \textbf{Durations and offsets} & \textbf{What is checked} \\
\midrule
MTL, STL & Positions of a timed word; time points of a signal & Numerical constants & All behaviors of a system model, or one signal \\
Parametric logics (PLTL, PSTL) & Positions of a trace; time points of a signal & Parameters: free variables with one value for the whole formula & Which parameter values make a system or a signal satisfy the formula \\
TPTL & Positions of a timed word & Variables that a freeze quantifier binds to the current time & Same as for MTL \\
HS & Intervals; tracks of a Kripke structure in model checking & Qualitative interval relations; metric variants exist & All initial tracks of a system model, under homogeneity \\
QEL (EL) & Time points in $\mathbb{Q}$ (in EL, points of a metric monoid) & Terms over variables bound by $\exists$ and $\forall$ inside the formula & One annotated record; existential queries admit rational witnesses \\
\bottomrule
\end{tabularx}
\end{table*}

\subsection{Query Systems and Temporal Logics}\label{subsec:related-logics}

\emph{Query systems.} Relational systems support temporal data through the period predicates of
SQL:2011~\citep{kulkarni2012temporal}, which adapt Allen's interval
relations~\citep{allen1983maintaining}. Clinical cohort-identification tools also provide temporal
query facilities~\citep{cohort1,cohort2,cohort3}. QEL complements these execution systems by
specifying the phenotype independently of its implementation. A criterion such as $\varphi$ may
be implemented by self-joins and endpoint inequalities; QEL supplies the semantics against which
that implementation can be checked. Comparing execution efficiency with hand-written queries is
a separate task (Section~\ref{subsec:scope}).

Temporal logics have been used to provide such specifications. They differ in how durations enter a formula and in
what a formula is checked against (Table~\ref{tab:logic-comparison}).

\emph{Constant bounds.} Linear temporal logic (LTL) can state that an event eventually occurs, but
not within what delay~\citep{pnueli1977temporal}. Metric temporal logic \citep[MTL;][]{metric} and
signal temporal logic \citep[STL;][]{stl1,stl2} attach numerical bounds to the temporal operators,
as in $\boldBox_{70}S$; the bounds are constants, fixed when the formula is written.

\emph{Parametric bounds.} Parametric temporal logics replace the constants by parameters. They
include parametric LTL \citep[PLTL;][]{alur2001parametric} and PROMPT-LTL
\citep{kupferman2009liveness}, parametric metric interval temporal logic
\citep{digiampaolo2015parametric}, and parametric STL
\citep[PSTL;][]{asarin2012parametric,jin2015mining}, and they build on parametric timed automata
\citep{alur1993parametric}. A parameter is a free variable with one value for the whole formula,
and the questions studied concern the set of values under which a system or a signal satisfies the
formula: whether it is empty or universal, and how to compute it. Several decidable fragments restrict each parameter to one polarity, yielding monotone
satisfaction~\citep{alur2001parametric,digiampaolo2015parametric}. QEL permits nested and alternating quantifiers within the formula. A variable may occur in
several displacements and windows, with a two-sorted quantifier scoping domain. The two
approaches meet on a single record: the values of the free variables under which a record
satisfies a QEL formula form a $\Q$-semilinear set (Proposition~\ref{prop:qel-semilinear}). This
is the counterpart of the validity domain that \citet{asarin2012parametric} compute for PSTL on a
single signal, also through quantifier elimination in linear arithmetic.

\emph{Quantified time and data.} Two further logics have quantifiers of a different kind. The
freeze quantifier of timed propositional temporal logic (TPTL) binds a variable to the time of the
current position, and constraints compare the bound times~\citep{alur1994really}. Metric
first-order temporal logic quantifies over the data values carried by events, while its time
bounds remain constants~\citep{basin2015monitoring}. QEL quantifies independently over rational offsets and positive durations.

\begin{table*}[t]
\caption{QEL in relation to earlier work on Ensemble Logic.}
\label{tab:prior-work}
\centering
\footnotesize
\begin{tabularx}{\textwidth}{@{}>{\raggedright\arraybackslash}p{0.11\textwidth}>{\raggedright\arraybackslash}p{0.17\textwidth}>{\raggedright\arraybackslash}p{0.29\textwidth}>{\raggedright\arraybackslash}X@{}}
\toprule
 & \textbf{TEL}~\citep{zhang2024temporal}
 & \textbf{$\EL(\Nat)$}~\citep{droste2026discrete}
 & \textbf{QEL (this paper)} \\
\midrule
Time domain
 & Positive ordered monoid; theory for $\Nat^{+}$
 & $\Nat$
 & $\Q$ (ordered group; signed displacement) \\
Models
 & Interpretations $\alpha\colon M^{+}\to 2^{\mathsf{P}}$
 & $\omega$-words; finite traces
 & LINE: finite interval ensembles (Definitions~\ref{interval}--\ref{tes-semantics}) \\
Main results
 & Proof system; undecidable satisfiability; expressiveness
 & $\Sigma^1_1$-complete satisfiability; expressiveness; relatively complete Hilbert system;
   NP-complete existential fragment
 & Morphological semantics and closure (Propositions~\ref{prop:ie-closure}
   and~\ref{closed}); finite rational evidence sets (Theorem~\ref{thm:finite-test-set});
   semi-algebraic closure for spatial supports (Theorem~\ref{thm:sa-closure}) \\
Model checking (single model)
 & Listed as open
 & Finite active-domain semantics: PTIME data, PSPACE-complete combined complexity
 & Endpoint arrays (Algorithms~\ref{alg:ensemble_ops}--\ref{alg:evidence}); polynomial data
   complexity in the number of endpoints (Theorem~\ref{complexity}) \\
Application
 & Outlook
 & ---
 & GTC and PGES cohorts on Center for SUDEP Research data; conformance assessment; Cohort
   Builder (Section~\ref{sec:application}) \\
\bottomrule
\end{tabularx}
\end{table*}

\emph{Interval logics.} Allen's interval algebra classifies the thirteen possible relations
between two intervals~\citep{allen1983maintaining}, and Halpern--Shoham logic (HS) turns these
relations into modal operators~\citep{halpern1991propositional}. Satisfiability is undecidable for
HS and for most of its fragments
\citep{halpern1991propositional,bresolin2011s,marcinkowski2011ultimate}. Model checking, by
contrast, is decidable~\citep{molinari2016checking} and has been studied extensively: HS is
interpreted over a finite Kripke structure, each finite path (track) of which is an interval,
under the homogeneity assumption that a proposition holds on an interval exactly when it holds at
each of its points. In this setting the expressiveness of HS has been compared with that of
point-based logics~\citep{bozzelli2019interval}; metric variants of interval logics also
exist~\citep{bresolin2013metric}. QEL shares the homogeneity idea but expresses it inside the
logic, through $\boldBox_t$. It is point-based and reaches intervals through windows. The problem
also differs: HS model checking asks whether every initial track of a system model satisfies a
formula, whereas this paper checks a formula against individual records, represented as  labeled collections 
of interval ensembles (LINE). 

\emph{What is a model and what is checked.} In the terminology of formal methods, evaluating a formula on one record
is path checking~\citep{markey2003model,bundala2014complexity} or offline
monitoring~\citep{leucker2009brief,basin2015monitoring}, here for a dense-time, first-order logic.
Verifying that every behavior of a system model satisfies a
formula~\citep{clarke1999model,baier2008principles} is a different problem
(Section~\ref{subsec:prior-el}). We keep the term model checking in its logical sense
(Section~\ref{intro}): each data record is a model, represented as a LINE.

\subsection{Earlier Work on Ensemble Logic}\label{subsec:prior-el}

The three operators, with first-order quantification over time, were introduced under the name
Temporal Ensemble Logic \citep[TEL;][]{zhang2024temporal}. TEL grew out of Temporal Cohort
Logic~\citep{TCL}, with a smaller and more portable operator basis. It defines the operators over
a positive ordered monoid and develops their theory for discrete time, including a proof system
and the undecidability of satisfiability. \citet{droste2026discrete} sharpen these results for the
discrete logic $\EL(\Nat)$, whose satisfiability problem is $\Sigma^1_1$-complete, and show that
finite active-domain evaluation has PTIME data complexity (Table~\ref{tab:prior-work}).

Neither paper develops the LINE algorithms for dense rational time studied here. 
TEL observes that dense and discrete settings ``are likely
to require different treatments with different results,'' and lists as open the data structures
and algorithms for logic-based cohort discovery, the complexity of model checking, reference to
the past, and interval and spatial enrichments~\citep{zhang2024temporal};
\citet{droste2026discrete} set aside the correspondence between biomedical data records and
semantic models. The present paper takes up these questions for rational time, the natural setting
of timestamped clinical annotations: contributions C2--C4 (Section~\ref{subsec:contributions})
clarifies the distinct contributions of this paper in these context (Table~\ref{tab:prior-work}). 
The input representations also differ: the discrete finite active-domain algorithm operates
on represented positions~\citep{droste2026discrete}, whereas our LINE-operation bounds
depend on annotation endpoints. 

Two standard problems affect  the limits of automation for EL. Satisfiability, the question whether
some model satisfies a formula, is undecidable already over discrete time~\citep{zhang2024temporal}, and with it the validity problem.
So is the verification of all behaviors of a finite-state system, because a system that can
produce every trace satisfies a formula exactly when the formula is valid. Rational time is no easier: it is
undecidable whether some record satisfies a given QEL formula (Proposition~\ref{prop:sat} in the
Appendix). Cohort discovery needs neither:
it evaluates a given formula on given finite records, a problem that is decidable with polynomial
data complexity (Theorem~\ref{complexity}).

\section{Ensemble Logic Over a Metric Monoid}
\label{metricmon}

We first recall EL~\citep{zhang2024temporal} over an arbitrary metric monoid, where points represent times
or spatial locations (Sections~\ref{syntax} and~\ref{subsec:el-semantics}), and then define its
rational-time variant, QEL (Section~\ref{subsec:qel}). These logical foundations precede the LINE
semantics and algorithmic results in Section~\ref{sec:qel-rational}.

\subsection{Syntax}
\label{syntax}

Displacement uses addition, whereas windows use distance. A \emph{metric monoid} is a monoid
$(M,+,0)$ equipped with a metric $d$ under which left and right translations are non-expanding.
We assume throughout the stronger condition of translation invariance: for all $a,b,e\in M$,
\[
d(a+e,\,b+e)=d(a,b)=d(e+a,\,e+b).
\]
Write $\|a\|:=d(a,0)$ for the \emph{norm} of $a$; the property of translation invariance gives
$d(a,b+a)=\|b\|$. Our principal domain is $\Q$, with $d(a,b)=|a-b|$ and $\|a\|=|a|$.
Euclidean spaces $\R^n$~\citep{rudin1976} and the integers provide corresponding domains for
spatial and discrete data.

EL has two sorts of terms, formed using addition and the norm. \emph{Point-terms} denote points
of $M$ and serve as displacements; \emph{norm-terms} have nonnegative real values and specify
window radii:
\[
u,v ::= a \mid x \mid u+v, \qquad s,t ::= c \mid z \mid \|u\| \mid s+t .
\]
Here $a\in M$ and $c>0$ are constants, $x$ is a \emph{point-variable}, and $z$ is a
\emph{norm-variable}. Norm-variables range over the positive reals. Each point-variable has a
fixed declared domain $D_x\subseteq M$, normally $M$ (restriction is stated explicitly).
A term without variables is \emph{closed} and is identified with its value. Every term used as a
modal radius must be strictly positive for all assignments in the declared domains. Thus
$\|x\|$ alone requires $D_x\subseteq M\setminus\{0\}$. 
Zero-radius modalities are excluded. With rational constants over $\Q$, closed terms can be
evaluated and compared exactly. Variable terms allow displacements and radii to vary within a
formula (Section~\ref{subsec:related-logics}).

Let $\mathsf{P}$ be a set of atomic propositions. The formulas of $\EL(M)$, the ensemble logic
over $M$, are defined by
\[
\varphi,\psi ::= p \mid \varphi_{u} \mid \neg\varphi \mid \varphi\wedge\psi \mid
\boldBox_s\varphi \mid \mdiamond_s\varphi \mid \exists x\,\varphi \mid \forall x\,\varphi ,
\]
where $p\in\mathsf{P}$, $u$ is a point-term, $s$ is a norm-term, and $x$ is a variable of either
sort: point-variable or norm-variable. Disjunction and implication are defined from $\neg$ and $\wedge$ as usual, quantifiers bind
variables under the standard rules of scope, and a formula without free variables is
\emph{closed}.

An atomic proposition names a set of points. The \emph{displacement} $\varphi_u$ asserts that
$\varphi$ holds at the point displaced by $u$ from the evaluation point. Bounded existence
$\mdiamond_s\varphi$ and bounded universality $\boldBox_s\varphi$ assert that $\varphi$ holds at
some point, respectively every point, at distance less than $s$; these points form the
\emph{window} of radius $s$. Displacement uses addition and the modalities use the metric;
$\|u\|$ converts a point-term to a norm-term. Quantifiers range over their declared domains, and Boolean connectives
have their usual meaning. For forward temporal windows (Section~\ref{subsec:qel}), the operators
read ``at,'' ``within,'' and ``throughout.''

\begin{remark}
Unlike first-order linear temporal logic~\citep{fltl}, EL has no predicate symbols taking
individual arguments. 
In other words, its only predicate symbols are the atomic propositions, which take no arguments.
 EL formulas, nevertheless, define predicates of the variables in their
terms: $\varphi_x$, for example, depends on $x$, which may be quantified.
\end{remark}

\subsection{Semantics}
\label{subsec:el-semantics}

An \emph{interpretation} is a map $\alpha\colon M\to 2^{\mathsf{P}}$; the atomic proposition $p$
is true at the point $a$ if $p\in\alpha(a)$. Satisfaction is defined for closed formulas, and
quantifiers are evaluated by substitution, which keeps variable assignments out of the notation:
$\varphi[x\mapsto b]$ is the formula obtained from $\varphi$ by replacing every free occurrence of
the variable $x$ by the constant $b$.

\begin{definition}[Satisfaction]\label{semantics}
Let $\alpha$ be an interpretation. For a point $a\in M$ and a closed formula $\varphi$, the
relation $(\alpha,a)\models\varphi$, written $a\models\varphi$ when $\alpha$ is fixed, is defined
as follows. Here $x$ has declared domain $D_x$, and $z$ is a norm-variable:
\[
\begin{array}{@{}l@{\ \ \text{iff}\ \ }l@{}}
a\models p & p\in\alpha(a);\\
a\models\neg\varphi & a\not\models\varphi;\\
a\models\varphi\wedge\psi & a\models\varphi\ \text{and}\ a\models\psi;\\
a\models\varphi_{u} & u+a\models\varphi;\\
a\models\mdiamond_s\varphi & b\models\varphi\ \text{for some}\ b\in M\ \text{with}\ d(a,b)<s;\\
a\models\boldBox_s\varphi & b\models\varphi\ \text{for all}\ b\in M\ \text{with}\ d(a,b)<s;\\
a\models\exists x\,\varphi & a\models\varphi[x\mapsto b]\ \text{for some}\ b\in D_x;\\
a\models\forall x\,\varphi & a\models\varphi[x\mapsto b]\ \text{for all}\ b\in D_x;\\
a\models\exists z\,\varphi & a\models\varphi[z\mapsto c]\ \text{for some real}\ c>0;\\
a\models\forall z\,\varphi & a\models\varphi[z\mapsto c]\ \text{for all real}\ c>0.
\end{array}
\]
\end{definition}

Displacement moves the evaluation point, and the modal operators quantify over its window, an open
ball that contains the point itself and grows with the radius. We write $\varphi[\mathbf{x}]$ for
a formula whose free variables are among $\mathbf{x}=(x^1,\ldots,x^n)$, and $\varphi[\mathbf{a}]$
for the closed formula obtained by substituting constants $\mathbf{a}=(a^1,\ldots,a^n)$ for them;
satisfaction of $\varphi[\mathbf{x}]$ refers to an assignment from the declared domains.

\begin{example}[Windows in the plane]\label{ex:plane}
Let $M=\R^2$ with the Euclidean distance, and let the atomic proposition $p$ be true exactly at
the points of a region $R$. The window of radius $s$ at a point $a$ is the open disc of radius $s$
centered at $a$. Hence $a\models\boldBox_s p$ iff this disc lies inside $R$, and
$a\models\mdiamond_s p$ iff it meets $R$. Composing the two operators,
$a\models\mdiamond_s\boldBox_s p$ iff $a$ lies in some open disc of radius $s$ that lies inside
$R$. Section~\ref{subsec:morphological-ops} elaborates this spatial reading
(Theorem~\ref{thm:sa-closure}).
\end{example}

\begin{definition}[Entailment and equivalence]\label{telentailment}
A formula $\varphi$ \emph{entails} a formula $\psi$, written $\varphi\models\psi$, if
$(\alpha,a)\models\varphi$ implies $(\alpha,a)\models\psi$ for all interpretations $\alpha$ and
all points $a\in M$; free variables range over their declared domains. The two formulas are \emph{equivalent}, written $\varphi\equiv\psi$, if each entails
the other. For a closed formula $\varphi$, we write $\semantics{\varphi}$ for the set
$\{(\alpha,a)\mid(\alpha,a)\models\varphi\}$.
\end{definition}

Displacement commutes with the Boolean connectives and composes by addition:
\[
\begin{array}{@{}l@{\qquad}l@{}}
\neg(\varphi_{u})\equiv(\neg\varphi)_{u}, &
(\varphi_{u})_{v}\equiv\varphi_{u+v},\\
(\varphi\wedge\psi)_{u}\equiv\varphi_{u}\wedge\psi_{u}, &
(\varphi\vee\psi)_{u}\equiv\varphi_{u}\vee\psi_{u},
\end{array}
\]
with $\varphi_{0}\equiv\varphi$, and $\varphi_{u}\equiv\varphi_{v}$ whenever $u=v$. The modal
operators distribute over disjunction and conjunction, respectively, and are monotone.

\begin{proposition}\label{tel-entailment}
For all formulas $\varphi,\psi$, point-terms $u,v$, and positive modal radii $s,t$:
\begin{enumerate}
\item $\mdiamond_s\varphi\equiv\varphi\vee\mdiamond_s\varphi$ and
      $\mdiamond_s(\varphi\vee\psi)\equiv\mdiamond_s\varphi\vee\mdiamond_s\psi$;
\item $\boldBox_t\varphi\equiv\varphi\wedge\boldBox_t\varphi$ and
      $\boldBox_t(\varphi\wedge\psi)\equiv\boldBox_t\varphi\wedge\boldBox_t\psi$;
\item $\varphi\models\mdiamond_s\varphi$ and $\boldBox_s\varphi\models\varphi$;
\item if $s\le t$, then $\mdiamond_s\varphi\models\mdiamond_t\varphi$ and
      $\boldBox_t\varphi\models\boldBox_s\varphi$;
\item if $\varphi\models\psi$, then $\varphi_{u}\models\psi_{u}$,
      $\mdiamond_s\varphi\models\mdiamond_s\psi$, and $\boldBox_t\varphi\models\boldBox_t\psi$;
\item if $\|u\|<\|v\|$, then $\varphi_{u}\models\mdiamond_{\|v\|}\varphi$ and
      $\boldBox_{\|v\|}\varphi\models\varphi_{u}$.
\end{enumerate}
\end{proposition}

\begin{proof}
Laws 1--5 follow by unfolding Definition~\ref{semantics}, using that a point belongs to each of
its windows (laws 1--3) and that a window grows with its radius (law~4). For law~6, suppose that
$a\models\varphi_{u}$, that is, $u+a\models\varphi$. By invariance, $d(a,u+a)=\|u\|<\|v\|$, so
that $u+a$ lies in the window of radius $\|v\|$ at $a$, and $a\models\mdiamond_{\|v\|}\varphi$. If
instead $a\models\boldBox_{\|v\|}\varphi$, then $\varphi$ holds throughout that window, in
particular at $u+a$, and $a\models\varphi_{u}$. Strict  inequality is due to open window used.
\end{proof}


The two modal operators are dual, $\neg\mdiamond_s\varphi\equiv\boldBox_s\neg\varphi$, as
are the two quantifiers. With De Morgan's laws and the laws for displacement, every
negation can therefore be moved inward until it applies to an atomic proposition. A
formula of this resulting shape is said to be in \emph{negation normal form}. 
(The negation-free formulas of Proposition~\ref{prop:monotone} are the formulas in negation normal form in which no
atomic proposition is negated.)

\subsection{Rational Ensemble Logic}
\label{subsec:qel}

We now move to rational time, the central focus of this paper. 
Rational timestamps represent annotated event boundaries exactly
(Section~\ref{intro}).
Rational points can moreover be added and subtracted, compared, and divided into equal parts;
a structure with these properties is a \emph{dense ordered divisible abelian group}
(Appendix~\ref{app:doag}).  The dense ordered divisible abelian group $(\Q,+,<)$ supports the finite
evidence sets of Section~\ref{subsec:qe-finite-evidence} and the algorithms of Section~\ref{mc}.

For temporal criteria such as ``within the next $t$ units,'' QEL uses the future half of a ball,
closed at the evaluation point and open at its far end, as TEL does~\citep{zhang2024temporal}.
Displacements remain signed: unlike TEL's nonnegative timeline, $\Q$ permits references to the
past.

\begin{definition}[QEL]\label{qelsemantics}
\emph{Rational Ensemble Logic}, written QEL or $\EL(\Q)$, has the formulas of $\EL(M)$ for $M=\Q$,
with rational constants. An interpretation is a map $\tau\colon\Q\to 2^{\mathsf{P}}$. For a time
point $r\in\Q$, a closed point-term $u$, and a closed positive radius term $t$, satisfaction
$(\tau,r)\models\varphi$, again written $r\models\varphi$, is defined by
\[
\begin{array}{@{}l@{\ \ \text{iff}\ \ }l@{}}
r\models p & p\in\tau(r);\\
r\models\varphi_{u} & r+u\models\varphi;\\
r\models\mdiamond_t\varphi & q\models\varphi\ \text{for some}\ q\in\Q\ \text{with}\ r\le q<r+t;\\
r\models\boldBox_t\varphi & q\models\varphi\ \text{for all}\ q\in\Q\ \text{with}\ r\le q<r+t.
\end{array}
\]
The clauses for the Boolean connectives and the quantifiers are those of
Definition~\ref{semantics}. Point-variables, here called \emph{rational variables}, have declared
domain $\Q$, $\Q_{>0}$, or $\Q\setminus\{0\}$, with $\Q$ the default; norm-variables range over
$\Q_{>0}$. These fixed domains are part of the variable declarations.
\end{definition}

Definition~\ref{qelsemantics} is the instantiation of Definition~\ref{semantics} for $M=\Q$ and $d(a,b)=|a-b|$, with
rational radii and one change of substance: the window at $r$ is not the ball $\{q\mid |q-r|<t\}$
but its future half $[r,r+t)$. In the rest of the paper, $\EL(\Q)$ denotes this future-directed
logic. Entailment, equivalence, and $\semantics{\varphi}$ are defined as in
Definition~\ref{telentailment}. The laws of Section~\ref{subsec:el-semantics} remain valid, with
one adjustment: the last law of Proposition~\ref{tel-entailment} holds for forward displacements,
in the form that $0\le u<t$ implies $\varphi_{u}\models\mdiamond_t\varphi$ and
$\boldBox_t\varphi\models\varphi_{u}$.

Over $\Q$ the norm is absolute value. Norm-variable quantifiers can be eliminated: replace $z$
by $\|x\|$ for a fresh rational variable declared to range over $\Q\setminus\{0\}$, and quantify
$x$ in place of $z$. This preserves both existential and universal quantification because absolute
value maps that domain onto $\Q_{>0}$. Henceforth we quantify only rational variables. Those with
domain $\Q_{>0}$ are \emph{duration variables}; they may occur in both displacements and radii,
and we write $t$ for $\|t\|$ in a radius. 
The domain of a variable does not depend on the values of other variables.

\begin{example}[The running example]\label{ex:record}
Let $\tau$ be the interpretation that the record of Figure~\ref{fig:gtc} determines: $S\in\tau(r)$
iff $10\le r<80$, $T\in\tau(r)$ iff $10\le r<30$, $C\in\tau(r)$ iff $30\le r<80$, and
$P\in\tau(r)$ iff $80\le r<120$. \emph{Windows.} We have $10\models\boldBox_{70}S$, because the
window $[10,80)$ lies within the seizure, but $10\not\models\boldBox_{71}S$, because the window
$[10,81)$ contains the time point $80$, at which $S$ is false. \emph{Quantifiers.} The formula
$\varphi$ of Section~\ref{subsec:operators} is closed; $x$ is a rational variable, and $a$ and $b$
are duration variables. It holds at the reference point, $0\models\varphi$: substituting the
witness $x=10$, $a=20$, $b=50$ gives the closed formula
$[\boldBox_{70}S\wedge\mdiamond_{20}T\wedge(\mdiamond_{50}C)_{20}]_{10}$, which holds at $0$
because its body holds at $10$. \emph{Signed displacement.} We have
$80\models\neg S\wedge S_{-1}$: the seizure is over, and it was still present one second earlier.
\end{example}

\begin{example}[Until]\label{ex:until}
Consider the following
\emph{until}  condition, whose terminating point lies strictly in the future, with the terminating formula written first:
$r\models\mathsf{U}(\varphi,\psi)$ iff some $q>r$ satisfies $\varphi$, while every $q'$ with
$r\le q'<q$ satisfies $\psi$. QEL expresses this condition as
\[
\exists t\,[\,\varphi_{t}\wedge\boldBox_t\psi\,],
\]
where $t$ is a fresh duration variable. Indeed, if $q$ is as required, put $t:=q-r$. Then
$r\models\varphi_{t}$ because $r+t=q$, and $r\models\boldBox_t\psi$ because the window $[r,r+t)$
is $[r,q)$. Conversely, a value of $t$ for which the body holds at $r$ yields $q:=r+t$. QEL thus expresses
this until operator over the rationals. 
The variable $t$ occurs in
a displacement and in a window, and this double use is what ties the time of $\varphi$ to the
extent of $\psi$.
\end{example}

Definition~\ref{qelsemantics} fixes the truth value of every closed formula at every time point.
Model checking, in the sense of this paper (Section~\ref{intro}), is the problem of deciding
whether $(\tau,q)\models\varphi$ for a given interpretation $\tau$, reference point $q$, and
closed formula $\varphi$. An interpretation specifies the true propositions at infinitely many
time points, so the problem presupposes that $\tau$ is finitely presented.
Section~\ref{sec:qel-rational} provides such a finite presentation: an annotated record determines its
interpretation through finitely many intervals, as in Example~\ref{ex:record}.

\section{Labeled Interval Ensembles}\label{sec:qel-rational}

A labeled interval ensemble assigns finitely many rational intervals to each atomic
proposition. It represents an annotated record as a QEL model without changing the satisfaction relation
(Sections~\ref{subsec:line} and~\ref{subsec:line-semantics}). Formula supports, the sets of time
points where formulas hold, can be obtained from interval endpoints
(Section~\ref{subsec:morphological-ops}), and the evaluation of quantification reduces to truth values on finite evidence sets
(Section~\ref{subsec:qe-finite-evidence}). These results provide the theory behind the algorithms of Section~\ref{mc}.

\subsection{Intervals, Ensembles, and LINEs}
\label{subsec:line}

Annotations mark intervals occupied by clinical events or states (Section~\ref{subsec:records}).
A LINE represents these annotations both as data records and as a semantic structure for QEL.

\begin{definition}\label{interval}
An \emph{interval} on $\Q$ is a non-empty set
$\{x\in\Q\mid a\,\lambda\,x\ \text{and}\ x\,\rho\,b\}$, where
$a\in\Q\cup\{-\infty\}$, $b\in\Q\cup\{\infty\}$, $a\le b$, and the \emph{boundary relations}
$\lambda,\rho$ are $<$ or $\le$. Inequalities at infinite endpoints are omitted. Write $[a,b]$,
$(a,b]$, $[a,b)$, or $(a,b)$, using square brackets for included finite endpoints and round
brackets otherwise, including at infinity. Singletons $\{a\}=[a,a]$  (abbreviated as $a$ in
interval lists) and $\Q=(-\infty,\infty)$ are intervals in this sense.
\end{definition}

\begin{definition}\label{ensemble}
An \emph{interval ensemble} $\delta$ is a finite set of mutually non-overlapping, that is, pairwise disjoint,
intervals; the empty set is an interval ensemble. We write $\IE$ for the set of all interval
ensembles over $\Q$, and $|\delta|:=\bigcup_{I\in\delta}I$ for the union of $\delta$. An interval
ensemble is \emph{canonical} if the union of any two distinct members is not again an interval.
\end{definition}

Thus $\{\;\}$ and $\{\Q\}$ are interval ensembles, but $\{\emptyset\}$ is not, because intervals are
assumed to be non-empty. The members of an interval ensemble may be adjacent, as $[1,2)$ and $[2,3]$ are, so
different ensembles can have the same union. Every finite union $A$ of intervals is, however, the
union of exactly one {\em canonical} interval ensemble, whose members are the maximal intervals contained
in $A$. Hence every interval ensemble $\delta$ has a \emph{canonical form}, the canonical ensemble
with union $|\delta|$, and an interval contained in $|\delta|$ lies in a single member of it.

\begin{example}\label{ex:ensembles}
The sets $\delta^1=\{(-\infty,0),\,(1,\infty)\}$ and $\delta^2=\{(1,2),\,[3,4),\,[4.1,5.2],\,6.5\}$
are canonical interval ensembles. The ensemble $\{[1,2),[2,3]\}$ is not canonical, and its canonical
form is $\{[1,3]\}$, whereas $\{(1,2),(2,3)\}$ is canonical, because $2$ lies in neither interval.
\end{example}

\begin{definition}\label{tes}
Let $\mathsf{P}$ be the set of atomic propositions, whose elements serve as the \emph{labels} of
annotations. A \emph{labeled interval ensemble} (LINE) is a (total) function
$\varepsilon\colon\mathsf{P}\to\IE$; it assigns to each label $\ell$ the ensemble
$\varepsilon(\ell)$ of the intervals that carry this label. The set of all LINEs is the function
space $[\mathsf{P}\to\IE]$. A pair $(\ell,\varepsilon(\ell))$ is called a \emph{temporal event},
labeled by $\ell$. A finite listing of non-empty temporal events specifies a LINE by assigning
$\emptyset$ to every unlisted label.
\end{definition}

\begin{example}\label{ex:line}
Over $\mathsf{P}=\{\ell^1,\ell^2,\ell^3\}$, and with $\delta^1$ and $\delta^2$ as in
Example~\ref{ex:ensembles}, the listings $\{(\ell^1,\delta^1),\,(\ell^2,\delta^2)\}$ and
$\{(\ell^3,\{\Q\})\}$ specify two LINEs; the second maps $\ell^1$ and $\ell^2$ to $\emptyset$. The
record of Figure~\ref{fig:gtc} is the LINE with the four temporal events $(S,\{[10,80)\})$,
$(T,\{[10,30)\})$, $(C,\{[30,80)\})$, and $(P,\{[80,120)\})$, consisting of singleton interval ensembles.
\end{example}

\subsection{LINE Semantics}
\label{subsec:line-semantics}

An interpretation $\tau\colon\Q\to 2^{\mathsf{P}}$ of QEL (Definition~\ref{qelsemantics}) is indexed
by time: $\tau(r)$ is the set of atomic propositions that are true at the time point $r$. A LINE is
indexed by labels: $|\varepsilon(p)|$ is the set of time points at which $p$ is true. Thus a LINE $\varepsilon$ determines the interpretation
\[
\tau_\varepsilon(r):=\{p\in\mathsf{P}\mid r\in|\varepsilon(p)|\},\qquad r\in\Q.
\]
For the record of Example~\ref{ex:line}, $\tau_\varepsilon$ is the interpretation $\tau$ of
Example~\ref{ex:record}. Not every interpretation arises in this way: those of the form
$\tau_\varepsilon$ are exactly the interpretations in which every atomic proposition is true on a
finite union of intervals. Two LINEs determine the same interpretation iff their corresponding ensembles have the same unions.

\begin{definition}\label{tes-semantics}
Let $\varepsilon$ be a LINE, $r\in\Q$ a time point, and $\varphi$ a closed formula of $\EL(\Q)$. We
write $(\varepsilon,r)\models\varphi$ if $(\tau_\varepsilon,r)\models\varphi$. Equivalently, the
relation $\models$ is defined by the clauses of Definition~\ref{qelsemantics}, with the base clause
in the form $r\models p$ iff $r\in|\varepsilon(p)|$, for $p\in\mathsf{P}$. The \emph{LINE semantics}
of $\varphi$ is the set
\[
\semantics{\varphi}^0:=\{(\varepsilon,q)\mid \varepsilon\in[\mathsf{P}\to\IE],\ q\in\Q,\
(\varepsilon,q)\models\varphi\}.
\]
\end{definition}

The notation $\semantics{\varphi}^0$ distinguishes this set from the set $\semantics{\varphi}$ of
Definition~\ref{telentailment}, which is taken over all interpretations:
$(\varepsilon,q)\in\semantics{\varphi}^0$ iff $(\tau_\varepsilon,q)\in\semantics{\varphi}$. Model checking therefore has a finite input: a closed formula $\varphi$, a reference point $q$,
and the temporal events of $\varepsilon$ whose labels occur in $\varphi$.

\subsection{Morphological Operations on Supports}
\label{subsec:morphological-ops}

Fix a LINE $\varepsilon$. The \emph{support} of a closed formula $\varphi$ in $\varepsilon$ is the
set of time points at which $\varphi$ is true,
\[
\semantics{\varphi}_{\varepsilon}:=\{q\in\Q\mid (\varepsilon,q)\models\varphi\},
\]
and model checking asks whether the reference point belongs to it. For an atomic proposition,
$\semantics{p}_{\varepsilon}=|\varepsilon(p)|$. The support of a compound formula arises from those
of its parts: by complement, intersection, and union for the Boolean connectives, and by the
following operations for displacement and the modal operators. They are the one-dimensional,
future-directed analogues of translation, dilation, and erosion in mathematical
morphology~\citep{matheron1975random,serra1982morphology,heijmans1994morphological}.

\begin{definition}[Future-directed morphological operations]\label{operations}
Let $A\subseteq\Q$, let $q\in\Q$, and let $r>0$ be rational. Put $H_r=[0,r)\cap\Q$. Define
\begin{align*}
A_q &:= \{x\in\Q\mid x+q\in A\},\\
\mdiamond_r A &:= \{x\in\Q\mid (x+H_r)\cap A\ne\emptyset\},\\
\boldBox_r A &:= \{x\in\Q\mid x+H_r\subseteq A\},\\
A^c &:= \Q\setminus A.
\end{align*}
Thus displacement is inverse translation, and $\mdiamond_r$ and $\boldBox_r$ 
are dilation and
erosion for the future window $x+H_r=[x,x+r)$, 
that is,
$\mdiamond_rA=A\oplus(-H_r)$ and $\boldBox_rA=A\ominus H_r$ in the notation of Lemma~2.
\end{definition}

\begin{lemma}[Directional morphological duality]\label{lem:directional-duality}
For every $A\subseteq\Q$ and every rational $r>0$,
\[
\boldBox_r A=(\mdiamond_r(A^c))^c.
\]
\end{lemma}

\begin{proof}
For $x\in\Q$, we have $x\in(\mdiamond_r(A^c))^c$ iff no point of the window $x+H_r$ belongs to
$A^c$, that is, iff $x+H_r\subseteq A$. This is the future-directed analogue of the standard duality
between erosion and dilation under complementation.
\end{proof}

\begin{proposition}[Closure of interval ensembles under temporal morphology]\label{prop:ie-closure}
Let $q\in\Q$ and let $r>0$ be rational; finite shifts leave infinite endpoints unchanged.
\begin{enumerate}
\item For an interval $I=\{x\in\Q\mid a\,\lambda\,x\ \text{and}\ x\,\rho\,b\}$, the operations are
computed by endpoint arithmetic:
\begin{align*}
I_q &= \{x\mid (a-q)\,\lambda\,x\ \text{and}\ x\,\rho\,(b-q)\},\\
\mdiamond_r I &= \{x\mid a-r < x\ \text{and}\ x\,\rho\,b\},\\
\boldBox_r I &= \{x\mid a\,\lambda\,x\ \text{and}\ x\le b-r\}.
\end{align*}
\item For a canonical interval ensemble $\delta$, the sets $|\delta|_q$, $\mdiamond_r|\delta|$, and
$\boldBox_r|\delta|$ are the unions of the sets $I_q$, $\mdiamond_r I$, and $\boldBox_r I$ over
$I\in\delta$, and $|\delta|^c$ consists of at most one interval before the first member of $\delta$,
one between any two consecutive members, and one after the last.
\end{enumerate}
Consequently, if $\delta$ is an interval ensemble, then $|\delta|_q$, $|\delta|^c$,
$\mdiamond_r|\delta|$, and $\boldBox_r|\delta|$ are finite unions of intervals.
\end{proposition}

\begin{proof}
(1) The rule for $I_q$ is clear. Let $W=x+H_r$. If $y\in W\cap I$, then $a\le y<x+r$ and
$x\le y\,\rho\,b$, so $a-r<x$ and $x\,\rho\,b$. Conversely, if $a-r<x$ and $x\,\rho\,b$, then
$W\cap I$ contains $x$ if $a\,\lambda\,x$, and otherwise $a$ or, if $a\notin I$, any rational $y$
with $a<y<\min(x+r,b)$. If $W\subseteq I$, then $a\,\lambda\,x$ because $x\in W$, and $x+r\le b$
because $W$ has rationals arbitrarily close to $x+r$; conversely, $a\,\lambda\,x$ and $x\le b-r$
give $a\,\lambda\,y$ and $y<b$ for $y\in W$.

(2) Displacement and dilation commute with unions. The window $x+H_r$ is an interval, so it lies
within $|\delta|$ iff it lies within one member of the canonical ensemble $\delta$
(Section~\ref{subsec:line}). The assertion on $|\delta|^c$ is clear, and the last assertion follows
by passing to the canonical form.
\end{proof}

\begin{figure}[pos=tp]
\centering
\begin{tikzpicture}[xscale=1, yscale=0.5]
\def\ptA{2.5}
\def\ptB{7}
\def\ptR{2}
\pgfmathsetmacro{\ptAR}{\ptA-\ptR}
\pgfmathsetmacro{\ptBR}{\ptB-\ptR}
\draw[->] (0,0) -- (8,0);
\draw[thick, blue] (\ptA,0.2) -- (\ptB,0.2);
\draw[blue] (\ptA,0.2) node[above] {$a$};
\draw[blue] (\ptB,0.2) node[above] {$b$};
\node[blue] at ({(\ptA+\ptB)/2},0.6) {$I$};
\draw[thick, red] (\ptAR,-0.2) -- (\ptBR,-0.2);
\draw[red] (\ptAR,-0.2) node[below] {$a-q$};
\draw[red] (\ptBR,-0.2) node[below] {$b-q$};
\node[red] at ({(\ptAR+\ptBR)/2},-0.7) {$I_q$};
\draw[->, dashed] (\ptA,0.2) .. controls (\ptA-0.5,1) .. (\ptAR,-0.2);
\draw[->, dashed] (\ptB,0.2) .. controls (\ptB-0.5,1) .. (\ptBR,-0.2);
\node at ({(\ptA+\ptAR)/2},0.9) {$q$};
\node at ({(\ptB+\ptBR)/2},0.9) {$q$};
\end{tikzpicture}
\begin{tikzpicture}[xscale=1, yscale=0.5]
\def\ptA{2.5}
\def\ptB{7}
\def\ptR{2}
\pgfmathsetmacro{\ptAR}{\ptA-\ptR}
\draw[->] (0,0) -- (8,0);
\draw[thick, blue] (\ptA,0.2) -- (\ptB,0.2);
\draw[blue] (\ptA,0.2) node[above] {$a$};
\draw[blue] (\ptB,0.2) node[above] {$b$};
\node[blue] at ({(\ptA+\ptB)/2},0.6) {$I$};
\draw[thick, red] (\ptAR,-0.2) -- (\ptB,-0.2);
\draw[red] (\ptAR,-0.2) node[below] {$a-r$};
\draw[red] (\ptB,-0.2) node[below] {$b$};
\node[red] at ({(\ptAR+\ptB)/2},-0.7) {$\mdiamond_r I$};
\draw[->, dashed] (\ptA,0.2) .. controls (\ptA-0.5,1) .. (\ptAR,-0.2);
\node at ({(\ptA+\ptAR)/2},0.9) {$r$};
\end{tikzpicture}
\begin{tikzpicture}[xscale=1, yscale=0.5]
\def\ptA{2.5}
\def\ptB{7}
\def\ptR{2}
\pgfmathsetmacro{\ptBR}{\ptB-\ptR}
\draw[->] (0,0) -- (8,0);
\draw[thick, blue] (\ptA,0.2) -- (\ptB,0.2);
\draw[blue] (\ptA,0.2) node[above] {$a$};
\draw[blue] (\ptB,0.2) node[above] {$b$};
\node[blue] at ({(\ptA+\ptB)/2},0.6) {$I$};
\draw[thick, red] (\ptA,-0.2) -- (\ptBR,-0.2);
\draw[red] (\ptA,-0.2) node[below] {$a$};
\draw[red] (\ptBR,-0.2) node[below] {$b-r$};
\node[red] at ({(\ptA+\ptBR)/2},-0.7) {$\boldBox_r I$};
\draw[->, dashed] (\ptB,0.2) .. controls (\ptB-0.5,1) .. (\ptBR,-0.2);
\node at ({(\ptB+\ptBR)/2},0.9) {$r$};
\end{tikzpicture}

\caption{Future-directed interval operations: displacement, dilation, and erosion. The original
interval $I$ is shown in blue; the transformed interval is shown in red.}
\label{fig:ieop_example}
\vspace{-1\baselineskip}
\end{figure}
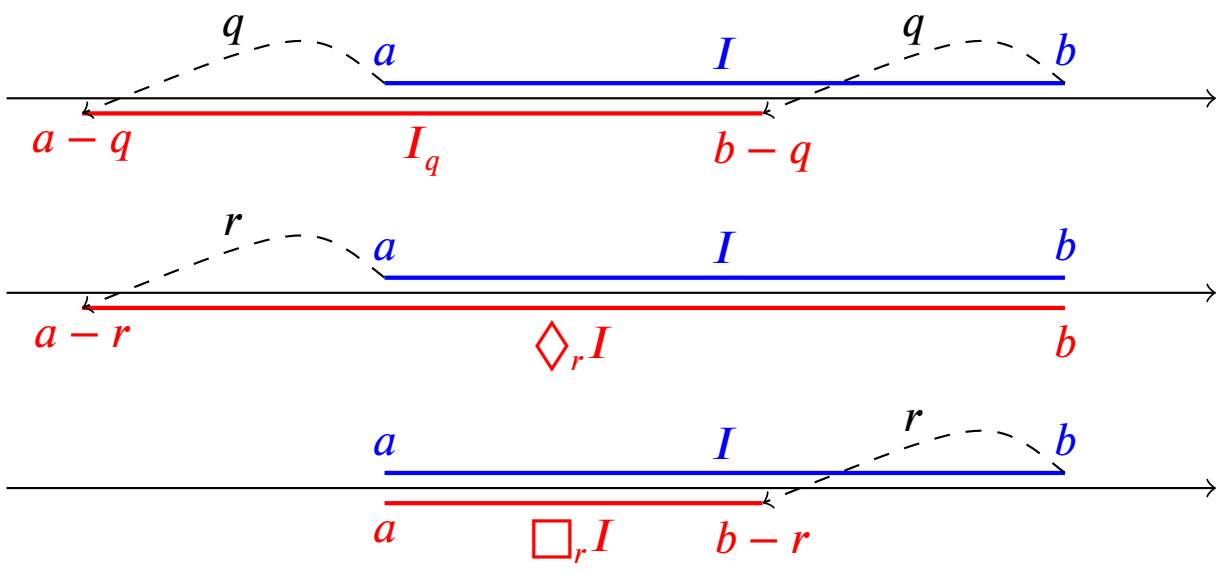

Part~(1) gives the geometric meaning of the modal operators (Figure~\ref{fig:ieop_example}).
Dilation extends $I$ into the past: its finite left boundary $a-r$ is open, and it inherits the
right boundary of $I$. Erosion inherits the left boundary and has a closed finite right boundary
at $b-r$. For the record of
Figure~\ref{fig:gtc}, $\mdiamond_{20}[10,30)=(-10,30)$ and $\boldBox_{70}[10,80)=[10,10]$ are the
supports of $\mdiamond_{20}T$ and $\boldBox_{70}S$ in Example~\ref{ex:record}. Eroded sets that are
empty are discarded; overlapping intervals, or adjacent intervals whose union is an interval,
are merged.

\begin{definition}[Lifting to interval ensembles and LINE]\label{event-op}
For an interval ensemble $\delta$, the operations $\delta_q$, $\mdiamond_r\delta$,
$\boldBox_r\delta$, and $\delta^c$ are defined to be the canonical interval ensembles whose unions
are respectively $|\delta|_q$, $\mdiamond_r|\delta|$, $\boldBox_r|\delta|$, and $|\delta|^c$, which
exist by Proposition~\ref{prop:ie-closure}; intersections and unions of interval ensembles are
understood in the same way. For a LINE $\varepsilon$ and an atomic proposition $p$, the support of
$p$ after an operation is obtained by applying the corresponding operation to $\varepsilon(p)$. For
a compound formula, the operations are applied to the support of the formula, not to the ensembles
of all labels of $\varepsilon$ simultaneously.
\end{definition}

\begin{definition}\label{operation-epsilon}
For a set $X\subseteq[\mathsf{P}\to\IE]\times\Q$ of pairs of a LINE and a time point, define
\begin{align*}
X^c &:= ([\mathsf{P}\to\IE]\times\Q)\setminus X,\\
X_q &:= \{(\varepsilon,x)\mid (\varepsilon,x+q)\in X\},\\
\mdiamond_r X &:= \{(\varepsilon,x)\mid \exists y\in x+H_r\;\;(\varepsilon,y)\in X\},\\
\boldBox_r X &:= \{(\varepsilon,x)\mid \forall y\in x+H_r\;\;(\varepsilon,y)\in X\}.
\end{align*}
These are semantic operations on truth sets of formulas: in a fixed LINE $\varepsilon$, they are the
operations of Definition~\ref{operations} on the set $\{x\mid(\varepsilon,x)\in X\}$, which for
$X=\semantics{\varphi}^0$ is the support $\semantics{\varphi}_{\varepsilon}$.
\end{definition}

\begin{proposition}[Compositional LINE semantics]\label{closed}
For closed QEL formulas $\varphi,\psi$, $q\in\Q$, and rational $r>0$, the following identities hold:
\begin{enumerate}
\item $\semantics{\varphi\wedge\psi}^0=\semantics{\varphi}^0\cap\semantics{\psi}^0$;
\item $\semantics{\varphi\vee\psi}^0=\semantics{\varphi}^0\cup\semantics{\psi}^0$;
\item $\semantics{\neg\varphi}^0=(\semantics{\varphi}^0)^c$ and
      $\semantics{\varphi_q}^0=(\semantics{\varphi}^0)_q$;
\item $\semantics{\mdiamond_r\varphi}^0=\mdiamond_r(\semantics{\varphi}^0)$ and
      $\semantics{\boldBox_r\varphi}^0=\boldBox_r(\semantics{\varphi}^0)$.
\end{enumerate}
\end{proposition}

\begin{proof}
The Boolean clauses are immediate from the satisfaction relation. For displacement,
$(\varepsilon,x)\models\varphi_q$ iff $(\varepsilon,x+q)\models\varphi$, which is
$(\varepsilon,x)\in(\semantics{\varphi}^0)_q$. For the diamond clause,
$(\varepsilon,x)\models\mdiamond_r\varphi$ iff $(\varepsilon,y)\models\varphi$ for some $y\in\Q$
with $x\le y<x+r$, which is membership in $\mdiamond_r(\semantics{\varphi}^0)$; the box clause is
dual. Each equality follows directly from QEL satisfaction.
\end{proof}

Proposition~\ref{closed} states that the LINE semantics is compositional. Read in a fixed
LINE $\varepsilon$, it says that the Boolean connectives act on supports by intersection,
union, and complement, and that displacement and the two modal operators act on them by
the operations of Definition~\ref{operations}. Model checking therefore needs a class of
sets that contains the supports of the atomic propositions and is closed under all of
these operations. Over $\Q$ the finite unions of intervals form such a class
(Proposition~\ref{prop:ie-closure}), so that the support of every quantifier-free formula
can be computed on endpoints (Corollary~\ref{cor:qf-support}). In the plane the finite
unions of disks do not: the intersection of two disks is a lens. The semi-algebraic sets
take their place (Theorem~\ref{thm:sa-closure}).


\begin{corollary}[Supports of quantifier-free formulas]\label{cor:qf-support}
Let $\varepsilon$ be a LINE and $\varphi$ a quantifier-free QEL formula, with fixed values of its
variables. The support $\semantics{\varphi}_{\varepsilon}$ is the union of an interval ensemble,
whose canonical form is obtained from the ensembles $\varepsilon(p)$ by the operations of
Definition~\ref{event-op} along the structure of $\varphi$.
\end{corollary}

\begin{proof}
By induction on $\varphi$, starting from $\semantics{p}_{\varepsilon}=|\varepsilon(p)|$. By
Proposition~\ref{closed}, read in the fixed LINE $\varepsilon$, each connective or operator acts on
supports by a Boolean operation or an operation of Definition~\ref{operations}, and finite unions of
intervals are closed under these by Proposition~\ref{prop:ie-closure}.
\end{proof}

Corollary~\ref{cor:qf-support} is the algorithmic content of this subsection: quantifier-free model
checking computes the canonical ensemble of the support by endpoint arithmetic and tests whether $q$
lies in one of its intervals.

\subsubsection{Spatial semi-algebraic closure}

For spatial data, Euclidean balls replace the temporal window $H_r=[0,r)$, and semi-algebraic
sets replace interval unions (Example~\ref{ex:plane}). The following closure result applies to
these spatial supports.

\begin{definition}[Semi-algebraic spatial supports]\label{def:semialgebraic}
A set $A\subseteq\R^n$ is \emph{semi-algebraic} if it is definable by a quantifier-free Boolean
combination of polynomial equalities and inequalities with real coefficients. Equivalently, by
quantifier elimination for real closed fields, $A$ is definable, with real parameters, by some
first-order formula in the language of ordered rings
$(0,1,+,-,\cdot,<)$~\citep{tarski1951decision,seidenberg1954new,bochnak1998real,basu2006algorithms,vandenDries1998tame}.
For $r\ge0$, let $B_r=\{z\in\R^n\mid \sum_{i=1}^n z_i^2\le r^2\}$.
We state the result for closed balls; replacing $\le$ by $<$ gives EL's open windows.
\end{definition}

\begin{lemma}[Euclidean morphological duality]\label{lem:euclidean-duality}
For every $A\subseteq\R^n$ and $r\ge0$,
\[
A\ominus B_r=(A^c\oplus B_r)^c,
\]
where $A\oplus B_r=\{a+b\mid a\in A,\ b\in B_r\}$ and $A\ominus B_r=\{x\mid x+B_r\subseteq A\}$.
\end{lemma}

\begin{proof}
For $x\in\R^n$, we have $x\in(A^c\oplus B_r)^c$ iff $x-b\in A$ for all $b\in B_r$. Because $B_r$ is
centrally symmetric, this holds iff $x+b\in A$ for all $b\in B_r$, that is, iff $x+B_r\subseteq A$.
\end{proof}

\begin{theorem}[Semi-algebraic closure under Euclidean morphology]\label{thm:sa-closure}
The class of semi-algebraic subsets of $\R^n$ is closed under complement, displacement by a fixed
vector, Euclidean dilation by $B_r$, and Euclidean erosion by $B_r$ for every fixed $r\ge0$.
Moreover, if the radius is treated as an additional variable constrained by $r\ge0$, the
corresponding sets of pairs $(x,r)$ are semi-algebraic.
\end{theorem}

\begin{proof} The proof is provided in Appendix~\ref{app:sa}.
\end{proof}

Rectangles, polygons, ellipses, disks, and their Euclidean rounded margins are semi-algebraic.
Theorem~\ref{thm:sa-closure}, whose proof also applies to open balls, shows that displacement and
both EL modalities preserve this class. It is a closure result for arbitrary real parameters;
effective computation additionally requires an effective coefficient domain, such as rational or
real-algebraic numbers.

\subsection{Quantifier Elimination and Evidence Sets}
\label{subsec:qe-finite-evidence}

Although QEL quantifiers range over infinitely many values, a fixed LINE admits finite test sets.
We first translate satisfaction into the first-order theory of $(\Q,+,<,0)$ and eliminate its
quantifiers (Proposition~\ref{prop:qel-semilinear}). The resulting linear conditions partition the
admissible assignments into finitely many polyhedral cells on which truth is constant. One rational
representative per cell then suffices for an existential or universal block
(Theorem~\ref{thm:finite-test-set}). These results justify
Algorithms~\ref{alg:model_checking} and~\ref{alg:evidence}.

The translation uses addition, order, and rational constants, but no products of variables. A
statement that is built from these with the quantifiers ``for some'' and ``for all,'' ranging over
time points, can always be restated without quantifiers: a time point $z$ with $x<z<y$, for
instance, exists exactly when $x<y$. This property is \emph{quantifier elimination} for the
first-order theory of dense ordered divisible abelian groups, a classical result that is commonly
attributed to~\citet{szmielew1955}; see \citet[Section~3.1]{marker2002} for a proof and
Appendix~\ref{app:doag} for the notions involved. It is the linear analogue of the
Tarski--Seidenberg theorem.


\begin{lemma}\label{lem:qe-doag}
\textup{[Quantifier Elimination for Dense Ordered Divisible Abelian Groups]}
Let $\mathcal{L}_{\mathrm{oag}} = \{+, -, <, 0, \{c\}_{c \in \Q}\}$ be the first-order language of dense
ordered abelian groups, with a constant for each rational number. The theory of
$(\Q, +, -, <, 0, \{c\}_{c \in \Q})$ admits quantifier elimination: every
$\mathcal{L}_{\mathrm{oag}}$ formula is equivalent to a quantifier-free Boolean combination of
atomic formulas of the form $\sum_i n_i x_i + c \sim 0$ with $n_i \in \mathbb{Z}$, $c \in \Q$, and
$\sim \in \{=, <\}$, which can be computed from the formula. Consequently, the definable subsets of
$\Q^n$ are the finite Boolean combinations of half-spaces, that is, the $\Q$-semilinear sets.
\end{lemma}

Integer coefficients abbreviate repeated addition and additive inverses, with $0x=0$; they do not
extend the language. Quantifier elimination already holds without named rational constants.
Treating the constants occurring in a formula as parameters, and clearing denominators, shows that
the resulting constants can be taken to be integer combinations of the original ones.

Fix a LINE $\varepsilon$. For a QEL formula $\varphi$, the finite \emph{threshold set}
$\Theta_{\varepsilon,\varphi}\subset\Q$ comprises the rational constants of $\varphi$ and the finite
endpoints of the intervals of $\varepsilon(p)$, for the atomic propositions $p$ of $\varphi$; for a
reference point $q$, we put $\Theta_{\varepsilon,q,\varphi}:=\Theta_{\varepsilon,\varphi}\cup\{q\}$.
Variables range over their declared domains (Section~\ref{syntax}). Write
$A(\mathbf{x})$ for the conjunction of the domain constraints on the free variables: $x>0$ for
durations and $x\ne0$ for nonzero rational variables. Unrestricted variables impose no constraint.
Modal radii are positive throughout these domains by well-formedness. These domain constraints
are included in the satisfaction sets and finite test sets below.

\begin{lemma}\label{lem:definability}
For every QEL formula $\varphi[\mathbf{x}]$ with free rational variables
$\mathbf{x} = (x^1, \ldots, x^n)$, there exists a quantifier-free
$\mathcal{L}_{\mathrm{oag}}$-formula $\Phi_\varphi(\mathbf{x}, y)$, whose constants are integer
combinations of the thresholds in $\Theta_{\varepsilon,\varphi}$, such that for every admissible assignment
$\mathbf{x}$ and every evaluation point $y\in\Q$,
\[
(\varepsilon, y) \models \varphi[\mathbf{x}] \quad\Longleftrightarrow\quad
\Q \models \Phi_\varphi(\mathbf{x}, y).
\]
\end{lemma}

\begin{proof}
Proceed by structural induction, applying Lemma~\ref{lem:qe-doag} whenever a clause introduces
quantifiers. For an atom $p$, let $\Phi_p(y)$ be the disjunction of the interval-membership
conditions $a\,\lambda\,y\wedge y\,\rho\,b$ for $\varepsilon(p)$, omitting infinite-endpoint
constraints. Boolean connectives translate homomorphically, and displacement substitutes $y+u$ for
$y$, with bound variables renamed to avoid capture.

Over $\Q$, a norm $\|u\|$ is $|u|$. Replace it by a fresh variable $v$ satisfying
$(u\ge0\land v=u)\lor(u<0\land v=-u)$. Conjoin this unique-value constraint to the affected
translation and bind $v$ existentially, inside every original binder for a variable of $u$.
This local construction preserves truth, including under negation and universal quantification,
and avoids expanding all sign choices. Lemma~\ref{lem:qe-doag} eliminates the auxiliary variables.

For the modal cases, assume that the induction hypothesis provides a quantifier-free formula
$\Phi_\psi(\mathbf{x},z)$ expressing satisfaction of $\psi$ at $z$. For a radius $t$, introduce the
intermediate formulas
\[
\widehat{\Phi}_{\mdiamond_t\psi}(\mathbf{x},y)
:=
\exists z\,
\bigl(
y\le z \land z<y+t
\land \Phi_\psi(\mathbf{x},z)
\bigr),
\]
and
\[
\widehat{\Phi}_{\boldBox_t\psi}(\mathbf{x},y)
:=
\forall z\,
\bigl(
(y\le z \land z<y+t)
\rightarrow \Phi_\psi(\mathbf{x},z)
\bigr),
\]
where $z$ is fresh. These clauses express satisfaction at some, respectively every, point of
$[y,y+t)$. Replacing norm subterms as above yields formulas of $\mathcal{L}_{\mathrm{oag}}$;
Lemma~\ref{lem:qe-doag} supplies equivalent quantifier-free formulas.

For an original quantifier binding $x$, let $G_x$ express its declared domain. Translate the
existential case as $\exists x\,(G_x\land\Phi_\psi)$ and the universal case as
$\forall x\,(G_x\rightarrow\Phi_\psi)$, then eliminate $x$. Here $G_x$ is $x>0$ for a duration,
$x\ne0$ for a nonzero variable, and true otherwise. One can check that every step preserves satisfaction on the
declared domains: the claim about constants follows from the remark after Lemma 3.
\end{proof}

\begin{proposition}\label{prop:qel-semilinear}
\textup{[QEL is semilinear over a fixed LINE]}
Let $\varepsilon$ be a LINE and let $\varphi[\mathbf{x}]$ be a QEL formula with free rational
variables $\mathbf{x}=(x^1,\ldots,x^n)$. Then the set
\[
S_{\varepsilon,\varphi}\;:=\;\bigl\{(\mathbf{x},y)\in\Q^{n+1}\mid A(\mathbf{x}),\ (\varepsilon,y)\models\varphi[\mathbf{x}]\bigr\}
\]
is $\Q$-semilinear; more precisely, it is a finite Boolean combination of half-spaces with
coefficients in $\mathbb{Z}$ and constants that are integer combinations of the thresholds in
$\Theta_{\varepsilon,\varphi}$. In particular, for $y=q$, the satisfaction set
$S_\varphi^{q}=\{\mathbf{x}\mid A(\mathbf{x}),\ (\varepsilon,q)\models\varphi[\mathbf{x}]\}$ is $\Q$-semilinear.
\end{proposition}
\begin{proof}
The set is defined by $A(\mathbf{x})\land\Phi_\varphi(\mathbf{x},y)$, using
Lemma~\ref{lem:definability}. Its atomic conditions are linear equalities and inequalities of the
stated form. Substituting $y=q$ gives the last assertion.
\end{proof}

For closed $\varphi$, the set $S_{\varepsilon,\varphi}$ is its support. Every $\Q$-semilinear
subset of $\Q$ is a finite union of intervals with rational or infinite endpoints, so the closure
assertion of Corollary~\ref{cor:qf-support} extends to quantified formulas. A closed phenotype can
therefore be stored as a new annotation label.

By Proposition~\ref{prop:qel-semilinear}, the set $S_\varphi^{q}$ is a Boolean combination of the
half-spaces $h_i(\mathbf{x})<0$ and the hyperplanes $h_i(\mathbf{x})=0$ for finitely many affine
polynomials $h_1,\ldots,h_k$, whose constants are integer combinations of the thresholds in
$\Theta_{\varepsilon,q,\varphi}$. Include also the boundary hyperplanes of $A$, and let $\mathcal{H}$ be the resulting set of $m$
proper hyperplanes; constant polynomials contribute no hyperplane. The
arrangement $\mathcal{A}(\mathcal{H})$ divides $\Q^n$ into finitely many cells, two points lying in
the same cell if each $h_i$ has the same sign at both: the open cells, on which no $h_i$ vanishes,
and the faces that bound them. The truth value of $\varphi$ depends on these signs only and is
therefore constant on every admissible cell. A finite evidence set, or test set, results from picking one
rational representative per cell of admissible values.

\begin{theorem}\label{thm:finite-test-set}
\textup{[Finite rational test set for quantified QEL formulas]}
Fix a LINE $\varepsilon$, a reference point $q\in\Q$, and a QEL formula $\varphi[\mathbf{x}]$ with
free rational variables $\mathbf{x}=(x^1,\ldots,x^n)$. There exists a finite set $F\subseteq\Q^n$ of admissible assignments,
computable from $(\varepsilon,q,\varphi)$, such that
\[
(\varepsilon,q)\models\exists\mathbf{x}\,\varphi[\mathbf{x}]
\;\iff\;
\exists\mathbf{a}\in F\;\;(\varepsilon,q)\models\varphi[\mathbf{a}],
\]
\[
(\varepsilon,q)\models\forall\mathbf{x}\,\varphi[\mathbf{x}]
\;\iff\;
\forall\mathbf{a}\in F\;\;(\varepsilon,q)\models\varphi[\mathbf{a}].
\]
Moreover, $F$ may be taken to consist of one representative of every admissible cell of
 $\mathcal{A}(\mathcal{H})$, of every dimension.
Then $|F|=O((m+1)^n)$, where $m=|\mathcal{H}|$. For a fixed
formula, $m$ is bounded by a polynomial in $N=1+|\Theta_{\varepsilon,q,\varphi}|$. If $\varphi$ is quantifier-free, $\mathcal{H}$ can be
chosen without quantifier elimination, as the hyperplanes~\eqref{eq:family} of the Appendix, which
compare the endpoints of $\varepsilon$, shifted by the displacements and radii of $\varphi$, with
$q$ and with one another; then $m=O(N^2)$. If, in addition, every modal operator is applied to a closed formula, the comparisons with $q$, together with
the domain and norm-sign boundaries, suffice: $m=O(N)$ and $|F|=O(N^n)$.
\end{theorem}
\begin{proof}
The proof is provided in Appendix~\ref{app:finite}.
\end{proof}

The theorem treats one block of similar quantifiers (all $\forall$s or all $\exists$s). Its matrix may contain further quantifiers.
Alternating quantification type requires successive blocks: choose representatives for the outer block using its full
matrix, then recompute inner test sets after substituting each outer assignment. A single set of
tuples cannot in general be reused across an alternating prefix. 
The linear bound covers the
specifications of Section~\ref{sec:application}, whose modalities are applied
to closed formulas. 
The quadratic bound cannot be improved for
general quantifier-free matrices (Appendix~\ref{app:finite}).

\section{Model-Checking Algorithm}
\label{mc}

Let $D=\{\varepsilon^1,\ldots,\varepsilon^m\}$ be a database of LINEs, each representing one
patient or recording session. We give two compatible evaluation procedures. Quantifier-free
formulas are evaluated by computing their supports recursively and testing membership of the
reference point. Arbitrary formulas are compiled into linear arithmetic and decided by quantifier
elimination; a single quantifier block also admits finite-evidence enumeration.

Support evaluation is analogous to the labeling method of model checking, which computes the
states satisfying each subformula~\citep{clarke1999model}. The finite cells for variable assignments
likewise resemble the regions used in timed-automaton verification~\citep{baier2008principles}.
Here the model is an annotated record; existential queries can return representative
assignments witnessing the specified pattern.

The formal results require well-formed LINEs. Their clinical interpretation additionally depends
on the accuracy and completeness of the annotations.
 
\begin{assumption}[Well-formed and complete records]\label{assump:best}
Every record $\varepsilon$ in the database $D$ is a \TES{}
(Definition~\ref{tes}) such that:
\begin{enumerate}
  \item each annotation label is mapped to a finite set of non-overlapping
        intervals with rational endpoints, on a common timeline within the
        record;
  \item interval endpoints, including whether each endpoint is open or
        closed, reflect the annotated onsets and offsets;
  \item annotations are complete for the labels occurring in the formula
        being checked, so that the absence of a label at a time point denotes
        the absence of the corresponding event.
\end{enumerate}
\end{assumption}
 
Condition~1 is sufficient for the formal results. Conditions~2 and~3 connect logical matches to
clinical events; in particular, Condition~3 supports the closed-world reading of negated labels.

A canonical ensemble is stored in parallel arrays of endpoints and closure marks, where $0$ denotes
open and $1$ denotes closed. For example, the interval ensemble
$\delta^2$ of Example~\ref{ex:ensembles} is stored as
\begin{alignat*}{9}
\delta^2.\text{boundary} &= [&&1,
 &&2,
 &&3,
 &&4,
 &&\tfrac{41}{10},
 &&\tfrac{26}{5},
 &&\tfrac{13}{2},
 &&\tfrac{13}{2}] \\
\delta^2.\text{closure\_mark} &= [&&0,
 &&0,
 &&1,
 &&0,
 &&1,
 &&1,
 &&1,
 &&1\ \ ] .
\end{alignat*}
The primitives implement Definition~\ref{operations}: displacement shifts both endpoints;
dilation moves the left endpoint and makes it open; erosion moves the finite right endpoint and
makes it closed. Infinite endpoints remain open. Complement constructs the intervening and exterior
gaps with reversed closure marks. Each primitive operates on a copy, preserving the input arrays,
and handles the empty and whole-line ensembles. Normalization removes empty intervals and merges
intervals whose union is an interval.

\renewcommand{\algorithmicrequire}{\textbf{Input:}}
\renewcommand{\algorithmicensure}{\textbf{Result:}}

\begin{algorithm}[h]
\caption{Primitive interval-ensemble operations}
\label{alg:ensemble_ops}
\begin{algorithmic}[1]
\REQUIRE Canonical interval ensembles $\delta,\delta'$, rational shift $q$, positive
rational radius $r$
\ENSURE Canonical interval ensemble
\STATE \textbf{Function} \textsc{Displacement}$(\delta,q)$
\STATE \quad subtract $q$ from every finite boundary of $\delta$; preserve closure marks
\STATE \quad \textbf{return} \textsc{Normalize}$(\delta)$
\STATE \textbf{Function} \textsc{Dilation}$(\delta,r)$
\STATE \quad \textbf{for each interval endpoint pair in $\delta$ do}
\STATE \quad \quad subtract $r$ from a finite left endpoint; set its closure mark to $0$
\STATE \quad \textbf{return} \textsc{Normalize}$(\delta)$
\STATE \textbf{Function} \textsc{Erosion}$(\delta,r)$
\STATE \quad \textbf{for each interval endpoint pair in $\delta$ do}
\STATE \quad \quad if the right endpoint is finite, subtract $r$ and set its closure mark to $1$
\STATE \quad \textbf{return} \textsc{Normalize}$(\delta)$
\STATE \textbf{Function} \textsc{Complement}$(\delta)$
\STATE \quad $\gamma\leftarrow$ the gaps before, between, and after the members of
$\delta$, with the closure mark of every finite endpoint reversed
\STATE \quad \textbf{return} \textsc{Normalize}$(\gamma)$
\STATE \textbf{Function} \textsc{Intersection}$(\delta,\delta')$,
\textsc{Union}$(\delta,\delta')$
\STATE \quad $\gamma\leftarrow$ the result of one merging pass over the two sorted
boundary arrays
\STATE \quad \textbf{return} \textsc{Normalize}$(\gamma)$

\end{algorithmic}
\end{algorithm}


For a fixed assignment $\eta$ of the free rational variables, let $E(\chi)$ denote the
canonical interval ensemble whose union is the support of the closed instance
$\chi[\eta]$ in $\varepsilon$. Quantifier-free evaluation is interval based:
\begin{align*}
E(p) &= \varepsilon(p),\\
E(\neg\chi) &=
   \textsc{Complement}(E(\chi)),\\
E(\chi\wedge\psi) &=
   E(\chi)\cap
   E(\psi),\\
E(\chi\vee\psi) &=
   E(\chi)\cup
   E(\psi),\\
E(\chi_u) &=
   \textsc{Displacement}(E(\chi),\eta(u)),\\
E(\mdiamond_t\chi) &=
   \textsc{Dilation}(E(\chi),\eta(t)),\\
E(\boldBox_t\chi) &=
   \textsc{Erosion}(E(\chi),\eta(t)).
\end{align*}

By Proposition~\ref{prop:ie-closure} and Definition~\ref{event-op}, every $E(\chi)$ is a
canonical interval ensemble, the ensemble $\varepsilon(p)$ being taken in canonical form.
The record $\varepsilon$ satisfies the quantifier-free formula $\varphi$ at reference
point $q$ exactly when $q\in|E(\varphi)|$. By Lemma 1, erosion could
also be computed as the complement of the dilation of the complement.

\paragraph{Translation.}
\textsc{TranslateToDOAG} implements the structural translation of
Lemma~\ref{lem:definability}. Atoms become disjunctions of interval-membership conditions;
Boolean connectives translate homomorphically; displacement substitutes $y+u$ for the evaluation
point; and the modalities introduce quantifiers over $[y,y+t)$. Original quantifiers retain their
scope and declared-domain guards. Absolute values use locally bound graph variables, as in the
lemma, so this compilation does not enumerate sign combinations.

The lemma eliminates quantifiers at each inductive step. The routine instead retains them,
returning a first-order formula. Algorithm~\ref{alg:model_checking} substitutes the reference point
and eliminates all remaining quantifiers. Both constructions preserve satisfaction; their outputs
need not be syntactically identical. Propositions~\ref{prop:ie-closure} and~\ref{closed}
independently justify the interval operations used for quantifier-free instances.

\begin{algorithm}[h]
\caption{Model checking for first-order QEL over one LINE (via DOAG translation)}
\label{alg:model_checking}
\begin{algorithmic}[1]
\REQUIRE LINE $\varepsilon$, closed QEL formula $\varphi$, reference point $q\in\mathbb Q$
\ENSURE \textbf{true} iff $(\varepsilon,q)\models\varphi$
\STATE $\widehat{\Phi}(y)\leftarrow
\textsc{TranslateToDOAG}(\varepsilon,\varphi,y)$
\COMMENT{Lemma~\ref{lem:definability}}
\STATE $\widehat{\Phi}_q\leftarrow\widehat{\Phi}(q)$
\STATE $\Psi\leftarrow
\textsc{QE}_{\mathrm{doag}}(\widehat{\Phi}_q)$
\COMMENT{Lemma~\ref{lem:qe-doag}}
\STATE \textbf{return}
\textsc{EvaluateLinearBooleanFormula}$(\Psi)$
\end{algorithmic}
\end{algorithm}

An outer existential or universal block can alternatively be decided on the finite test set of
Theorem~\ref{thm:finite-test-set}, constructed by Algorithm~\ref{alg:evidence}. For a quantifier-free
matrix, evaluate each representative by the support equations. If the matrix contains quantifiers,
evaluate the substituted closed instance by Algorithm~\ref{alg:model_checking}.

For cohort discovery over a database $D=\{\varepsilon^1,\ldots,\varepsilon^m\}$, Algorithm~\ref{alg:model_checking}
is run on each record, returning
$
\{\varepsilon\in D\mid (\varepsilon,q)\models\varphi\}.
$
The evidence set depends on $(\varepsilon,q,\varphi)$ and must be valid for the record being
checked. Implementations may cache shared endpoint transformations without changing this requirement.

When the formula has a single outer existential or universal block, the quantifier-elimination step
can be implemented as a finite-evidence enumeration. Algorithm~\ref{alg:evidence} makes explicit what
the routine often called \textsc{critical\_points} must compute.

\begin{algorithm}[h]
\caption{Finite evidence set for one quantifier block}
\label{alg:evidence}
\begin{algorithmic}[1]
\REQUIRE LINE $\varepsilon$, reference point $q$, formula $\varphi[\mathbf{x}]$
\ENSURE Finite $F\subseteq\mathbb Q^n$ satisfying Theorem~\ref{thm:finite-test-set}
\STATE $\Phi(\mathbf{x})\leftarrow \textsc{TranslateToDOAG}(\varepsilon,\varphi[\mathbf{x}],q)$
\STATE $\Psi(\mathbf{x})\leftarrow \textsc{QE}_{\mathrm{doag}}(\Phi(\mathbf{x}))$
\STATE $A(\mathbf{x})\leftarrow$ declared free-variable domain constraints
\STATE $\mathcal H\leftarrow$ all proper affine boundary hyperplanes in $\Psi$ and $A$
\STATE compute the sign cells of $\mathbb Q^n$ induced by $\mathcal H$, including all faces
\STATE choose one rational point from each nonempty cell satisfying $A$; collect these in $F$
\STATE \textbf{return} $F$
\end{algorithmic}
\end{algorithm}

For an existential block, stop at a satisfying representative or collect all satisfying
representatives as evidence; for a universal block, check every representative. With one variable,
the cells are boundary points, intervening open intervals, and exterior rays; retain only those in
the declared domain. With no hyperplanes, there is one cell. For a quantifier-free matrix, the
explicit family in Theorem~\ref{thm:finite-test-set} supplies $\mathcal H$ directly, avoiding
quantifier elimination. Alternating prefixes require the successive-block construction described
after that theorem, or the general procedure of Algorithm~\ref{alg:model_checking}.

A LINE is a total function: an omitted label denotes the empty ensemble. Its negation therefore
has whole-line support, as required by the closed-world semantics.

\subsection{Complexity Analysis}
\label{subsec:complexity}
\begin{theorem} \textup{[Correctness and data complexity]}\label{complexity}
Let $\varepsilon$ be a LINE, let $N$ be its number of stored boundary entries plus $2$, and let
$q\in\Q$ and $\psi$ be a reference point and a closed QEL formula. Count arithmetic operations and
comparisons on rational numbers.
\begin{enumerate}
\item Algorithm~\ref{alg:model_checking} returns \textbf{true} iff $(\varepsilon,q)\models\psi$.
  For fixed $\psi$, and with quantifiers eliminated by test terms, it runs in time polynomial
  in $N$.
\item Let $\psi$ be $\exists\mathbf{x}\,\chi$ or $\forall\mathbf{x}\,\chi$, where $\chi$ is
  quantifier-free and $\mathbf{x}=(x^1,\ldots,x^k)$ with $k\ge1$, and let $F$ be the evidence set of
  Algorithm~\ref{alg:evidence} for the explicit hyperplanes of Theorem~\ref{thm:finite-test-set}.
  Evaluating $\chi[\mathbf{a}]$ at $q$ by the support equations, for $\mathbf{a}\in F$, decides
  whether $(\varepsilon,q)\models\psi$. In the existential case, each satisfying representative
  can be returned as evidence. For fixed $\psi$, this takes time $\mathcal{O}(N^{2k+1})$, and time
  $\mathcal{O}(N^{k+1})$ if each modal operator of $\chi$ is applied to a closed formula.
\end{enumerate}
For record-local evaluation over a database $D$, the same bounds hold with $N$ equal to the total
number of stored boundary entries plus $2|D|$. Thus data complexity is polynomial for each fixed
formula, including databases containing empty records.
\end{theorem}

\begin{proof}
The proof is provided in Appendix~\ref{app:complexity}.
\end{proof}

Fixing the formula reflects cohort discovery, where a small specification is evaluated against a
large dataset~\citep{vardi1982complexity}. The GTC and PGES formulas of
Section~\ref{sec:application} satisfy the $\mathcal{O}(N^{k+1})$ bound, with $k=3$ and $k=4$,
respectively. With the formula also part of the input, model checking is PSPACE-hard and decidable
in exponential space (Proposition~\ref{prop:combined}).

Section~\ref{sec:application} reports prototype evaluations and a cohort-query interface. The
released checker searches annotation-boundary candidates; it does not implement the complete cell
construction described above. Its conformance results therefore concern the reported test cases,
while the general correctness guarantee applies to Algorithms~\ref{alg:model_checking}
and~\ref{alg:evidence}.

Empirical comparison of execution strategies and clinical validation of phenotype definitions are outside the scope of this paper; see Section~\ref{subsec:scope}.

\section{Application in Cohort Discovery}\label{sec:application}

We illustrate how QEL can be used for cohort discovery using the examples of generalized tonic-clonic (GTC) seizures and postictal
generalized EEG suppression (PGES). We give the temporal specifications, describe their evaluation
over annotated records, report a synthetic conformance assessment, and present the Cohort Builder
interface.

\subsection{Data Models and Cohorts}

Electrophysiological cohort discovery identifies recording sessions with specified temporal
patterns. Earlier work~\citep{TCL} formulated this task as temporal querying and evaluated it on
Center for SUDEP Research (CSR) EEG data~\citep{csrpaper,csrweb}. Those results are different from
the work reported here.

 \emph{SeizureSphere} is a graphical user interface that supports the authoring of structured queries about the ordering, duration, and
co-occurrence of annotated events. A user can select event types, such as tonic activity, clonic
activity, and EEG suppression, and specify temporal constraints through parameterized windows.
These constraints are translated into logical formulas, which are then systematically translated into database queries
 and evaluated over the annotated data records. The
formula states the query's meaning independently of the interface, interpretation of the query constructs and query execution strategy.

Each record represents annotated data set from a recording session, such as an Epilepsy Monitoring Unit (EMU) EEG
session, as a LINE $\varepsilon\colon\mathsf P\to\IE$. For each label it supplies a finite ensemble
of pairwise disjoint intervals on the rational timeline. Multiple sessions for one patient
may be represented by separate LINEs or a session-indexed family. Given a closed QEL formula
$\varphi$, the record-level cohort consists of the LINEs satisfying $\varphi$ at the reference
point.

\subsection{Logical Specification of GTC Seizures}
\label{gtc-qel}

The following formula expresses an ordered tonic-clonic pattern within a sustained seizure
annotation:
\begin{center}
$\varphi:= \exists x \exists a\exists b [ \boldBox_{a+b}S \wedge \mdiamond_{a}T \wedge
(\mdiamond_{b}C)_{a}]_{x}$,
\end{center}
where $x\in\Q$ and $a,b>0$ are duration variables. At reference point $0$, a witness requires
$S$ throughout $[x,x+a+b)$, some $T$ point in $[x,x+a)$, and some $C$ point in
$[x+a,x+a+b)$. Thus the chosen tonic occurrence precedes the chosen clonic occurrence, and both
lie within a continuous $S$ interval. At a general reference point $q$, all these endpoints are
translated by $q$.

The formula requires occurrences of $T$ and $C$, rather than continuous tonic and clonic phases.
It does not require their complete annotations to be disjoint, or identify $x$ and $x+a+b$ with
the onset and offset of a maximal seizure annotation. It specifies a temporal pattern compatible
with the intended phenotype, whose clinical interpretation depends on the labels and any
additional duration or boundary requirements.

\subsection{Logical Specification of PGES}
\label{subsec:pges}

PGES is a period of generalized EEG suppression after a seizure and has been studied as a possible
SUDEP risk marker~\citep{lhatoo2010electroclinical}. Its temporal relationship to seizure activity
motivates the illustrative specification
\begin{center}
$\psi:= \exists x \exists a\exists b\exists c\,
[\, \boldBox_{a+b}S \wedge \mdiamond_{a}T \wedge (\mdiamond_{b}C)_{a} \wedge
(\boldBox_{c}P)_{a+b}\,]_{x}$,
\end{center}
where $a,b,c>0$. The added conjunct requires $P$ throughout $[x+a+b,x+a+b+c)$ at reference point
$0$, immediately after the selected $S$ window. The shared displacement $a+b$ makes the two
windows adjacent.

This base specification is deliberately permissive: it does not require $S$ to cease at
$x+a+b$, prohibit earlier $P$ activity, or impose a clinical minimum suppression duration.
For example, $S=[0,100)$, $T=[0,1)$, $C=[1,2)$, and $P=[2,3)$ satisfy $\psi$ with
$(x,a,b,c)=(0,1,1,1)$ although seizure activity continues. For half-open seizure annotations,
adding $(\neg S)_{a+b}$ to the bracketed body forces the shared boundary to be an $S$ offset.
A prescribed minimum suppression duration $d>0$ can be imposed by adding $(\boldBox_dP)_{a+b}$.
The experiments below concern the displayed base formula and the stricter query defined next.

\subsection{Model-Checking and Cohort Identification}
\label{subsec:cohort-generation}

Let $D=\{\varepsilon^1,\ldots,\varepsilon^m\}$ be a database of LINE records. Its cohort for
$\psi$ is
\[
\semantics{\psi}_D:=\{\varepsilon\in D\mid(\varepsilon,q)\models\psi\}.
\]
Because the anchor $x$ ranges over all of $\Q$, membership is independent of $q$: a change of
reference point is absorbed by the witness for $x$. The model-checking procedure of
Section~\ref{mc} evaluates $\psi$ separately on each record, interpreting its atomic propositions
by the recorded annotations.

All modalities in the body of $\psi$ apply directly to atomic propositions. Its truth is therefore
determined by the positions of $q+x$, $q+x+a$, $q+x+a+b$, and $q+x+a+b+c$ relative to the
annotation endpoints, together with $a,b,c>0$. The relevant hyperplanes describe equality with
those endpoints and zero durations. The evidence-set construction samples the admissible sign
cells, including their lower-dimensional faces, so that boundary cases are retained.

A record qualifies exactly when some consistent assignment to its anchor and duration variables
satisfies $\psi$. The exact procedure obtains witnesses from representative cells. The prototype
returns the candidate matches and displays their annotation patterns. 
Re-evaluation of the same formula over
the same annotations yields the same record-level cohort. Clinical validity is beyond the scope because it additionally depends
on the correctness of phenotype definition and the annotation of data (Section~\ref{subsec:scope}).

\subsection{Synthetic Conformance Experiment}
\label{sec:synthetic-evaluation}

We manually constructed 54 records with seizure ($S$), tonic ($T$), clonic ($C$), and EEG
suppression ($P$) annotations on a 0 - 240\,s timeline. The cases cover temporal and duration
variation, multiple episodes, boundary configurations, intended query contrasts, and labeling or
relation errors. Expected outcomes were assigned from each case's designed structure. Besides
$\psi$, the released evaluation uses the following stricter formula, with
$N:=\neg(S\vee T\vee C\vee P)$ and $a,b,c>0$:
\[
\begin{aligned}
\psi_{\mathrm{str}}:=\exists x\exists a\exists b\exists c\,[\,&\boldBox_{10}N\;\wedge\\[-2pt]
 &(\boldBox_{a+b}(S\wedge\neg P))_{10}\;\wedge\\[-2pt]
 &(\mdiamond_a(T\wedge\neg C))_{10}\;\wedge\\[-2pt]
 &(\mdiamond_b(C\wedge\neg T))_{10+a}\;\wedge\\[-2pt]
 &(\boldBox_c(P\wedge\neg S))_{10+a+b}\,]_x .
\end{aligned}
\]
Here $x$ anchors a 10-second quiet prefix; seizure activity starts at $x+10$ when the reference
point is $0$. The tonic and clonic exclusions apply at their respective witness points, so their
full annotations may overlap. The dataset and checker are available at
\url{https://github.com/yhc29/QEL-demo}. The released base evaluation treats input intervals as
closed; the stricter evaluation converts them to half-open intervals. These conventions are part
of the experimental specification.

The base query matched 43 records and produced 47 patterns; the stricter query matched 33 records
and produced 36 patterns, retaining at most one witness per seizure component. Both agreed with
all 54 manually assigned record-level outcomes. Every
strict-positive case was base-positive; the stricter query rejected the 10 designed contrast cases
accepted by the base query, and both rejected the 11 designed labeling or temporal-relation
errors. These results document agreement on the constructed cases. Since the cases and expected
outcomes were designed together, this assessment neither proves implementation correctness nor
estimates clinical sensitivity or specificity. The checker searches observed endpoint candidates
and uses floating-point tolerances; it does not implement the full sign-cell enumeration of
Algorithm~\ref{alg:evidence}. Its agreement on these cases therefore does not establish complete
or exact evaluation for arbitrary LINE records.

\subsection{Cohort Builder Interface}
\label{subsec:builder}

\begin{figure*}[ht]
\centering
\IfFileExists{interface_1.pdf}{\includegraphics[width=\textwidth]{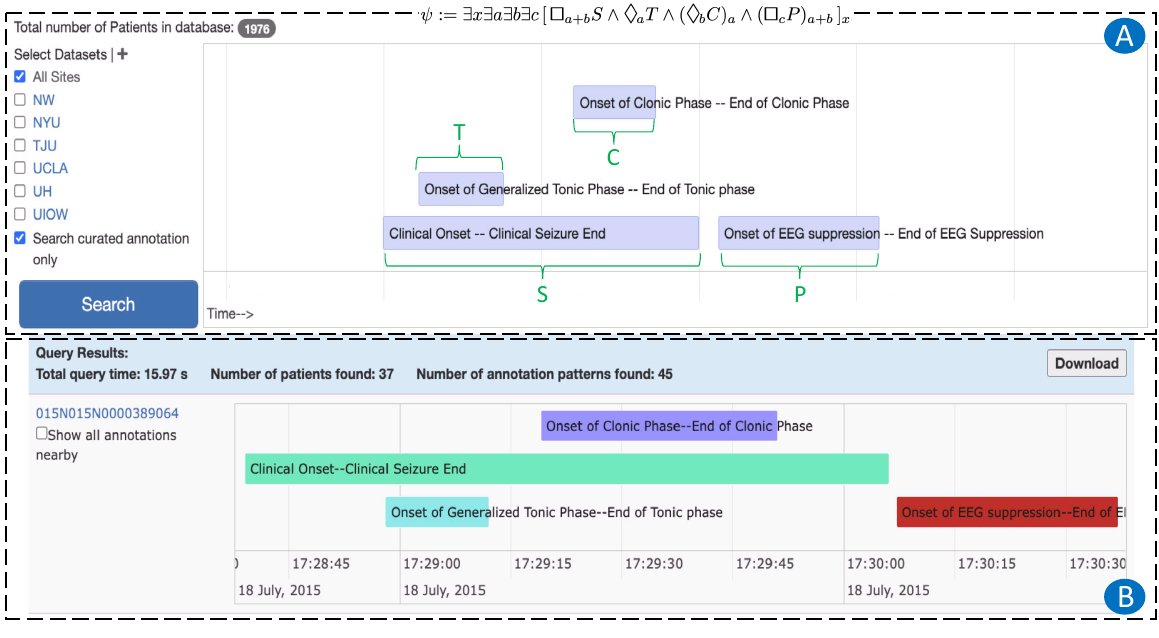}}{\fbox{\parbox{0.9\textwidth}{\centering Placeholder for the QEL-based Cohort Builder interface. Place \texttt{interface\_1.pdf} in the compilation directory.}}}
\caption{The QEL-based Cohort Builder interface, illustrated with the PGES specification $\psi$.
(A) The interactive Cohort Builder, where a query is constructed by composing clinical events along
a timeline. Each interval represents a clinical event: the generalized tonic phase ($T$), the clonic
phase ($C$), the overall clinical seizure ($S$), and the EEG suppression phase ($P$), and the
corresponding QEL formula is synthesized automatically and displayed at the top.
(B) The Query Results panel, which returns the list of patients satisfying the query and visualizes
the matched annotation pattern for each patient on a time-aligned axis.}
\label{fig:interface}
\end{figure*}

The \emph{Cohort Builder} presents temporal query construction through a timeline
(Figure~\ref{fig:interface}A). Researchers arrange clinical events and their temporal relationships;
the interface displays the corresponding QEL formula. In the example, the timeline represents
tonic activity ($T$), clonic activity ($C$), the seizure ($S$), and EEG suppression ($P$).
The formula determines aspects of the arrangement that are enforced. In $\psi$, for example,
the tonic and clonic conditions require occurrences within their windows, whereas the seizure
and suppression conditions require persistence. Data-source filters restrict evaluation to
selected annotation collections.

The Query Results panel (Figure~\ref{fig:interface}B) reports evaluation time, the number of
qualifying patients, and the number of matched annotation patterns. For each patient it displays
the constituent annotations as colored intervals on a common time axis. Researchers can inspect
the recorded timing and the evidence for inclusion, then export the cohort for further analysis.
The interface thus exposes both the logical specification and its matching annotations.

In the illustrated run, the database contained 1,976 patients. The PGES query returned 37 patients
and 45 distinct annotation patterns in 15.97 seconds. This is an illustrative feasibility result
for the query workflow; no comparative performance benchmark or clinical adjudication of the
returned cohort was conducted.

\subsection{Scope, Assumptions and Robustness}
\label{subsec:scope}

For well-formed LINE records, the procedures in Section~\ref{mc} compute logical satisfaction
exactly (Propositions~\ref{prop:ie-closure} and~\ref{closed};
Theorems~\ref{thm:finite-test-set} and~\ref{complexity}). This is a mathematical guarantee about
the specified procedures. It does not establish the relative efficiency of indexing or query
planning strategies, or the clinical validity of a phenotype definition. The synthetic assessment
checks agreement with expected outcomes on selected cases; the clinical cohort was not
independently adjudicated, and no false-discovery rate is reported. The released endpoint search
can miss satisfying assignments: for closed $S=T=C=[0,1]$ and $P=[1,2]$, $\psi$ holds with
$(x,a,b,c)=(0,\tfrac12,\tfrac12,1)$, but neither released search path for $\psi$ finds a match. The required split
lies between observed endpoints. Complete evaluation requires the representatives specified by
Theorem~\ref{thm:finite-test-set}, including representatives within open cells.

Annotations may be missing or spurious, contrary to Condition~3 of
Assumption~\ref{assump:best}. For formulas without negation, the following monotonicity result
determines the effect of these errors.

\begin{proposition}[Monotonicity of negation-free formulas]
\label{prop:monotone}
For LINEs $\varepsilon,\varepsilon'$ over the same label set $\mathsf P$, write
$\varepsilon\sqsubseteq\varepsilon'$ if $|\varepsilon(p)|\subseteq|\varepsilon'(p)|$ for every
$p\in\mathsf P$. Let $\varphi$ be built from atoms by conjunction, disjunction, displacement,
bounded modalities, and quantifiers. If $\varepsilon\sqsubseteq\varepsilon'$, then, at every
$q\in\Q$ and every admissible assignment of its free variables,
$(\varepsilon,q)\models\varphi$ implies $(\varepsilon',q)\models\varphi$.
\end{proposition}

\begin{proof}
Induct on $\varphi$, simultaneously for all admissible assignments. The atomic case is the
definition of $\sqsubseteq$; conjunction and disjunction preserve implication. If $A\subseteq B$,
then $A_u\subseteq B_u$, every window meeting $A$ meets $B$, and every window contained in $A$
is contained in $B$. Hence displacement and both modalities preserve inclusion of supports.
For either quantifier, apply the induction hypothesis to every instance in its declared domain,
which is the same in both records. Existential and universal quantification then preserve the
implication.
\end{proof}

The GTC and base PGES formulas contain no negation. For these formulas, missing annotations alone
can only remove records from the cohort, and spurious annotations alone can only add records.
When both errors occur, neither inclusion follows. Formulas containing negation or implication
need not be monotone, although some are; the stricter PGES query's exclusion conditions therefore
require separate analysis. Under the closed-world semantics, a missing annotation can make a
negated label true.

A LINE records intervals independently of how their annotations were produced. They may come
from expert curation or automated detectors, including those developed for
PGES~\citep{theeranaew2018,li2020pges}. The semantics interprets rational endpoints and their
closure marks exactly. Annotation resolution, endpoint conventions, and accuracy determine how
these results relate to the underlying recording. In particular, unrestricted signed
displacement can examine times outside the recorded session; negated labels there follow the
same closed-world convention. Queries intended to remain within an observed session must impose
that restriction, for example through an explicit observation-interval label.
Probabilistic or confidence-weighted annotations require an extension of the semantics.

\subsection{Generalizability}
\label{sec:broader-applications}

Related studies apply the temporal reasoning principles underlying QEL to other biomedical
representations. EL operators have been used to formalize narrative rules for arousals,
respiratory events, N2 sleep staging, and limb movements from the AASM Scoring Manual, making
durations, cross-signal relationships, and response latencies explicit~\citep{sleep}. TEL, which
shares EL's foundations but differs from the rational-time QEL developed here, has been applied
to clinical-trial protocols to express dependencies among clinical concepts and temporal
events~\citep{time25,scaleup}. QEL has also been applied to annotated sleep records from the
National Sleep Research Resource, represented as LINEs for temporal pattern
matching~\citep{huang_qel_nsrr_engine}.

These studies suggest that formal temporal specifications can be reused across narrative
guidelines, trial protocols, and timestamped observations.  In each case, reproducibility is a part of 
the progress toward TRUER biomedical knowledge representation and discovery, which
does not itself establish the validity of the data annotations nor the concordance of the specification formulas with the 
intended clinical content.

\section{Conclusion}\label{sec:conclusion}

QEL provides a formal language for temporal phenotypes in real-world electrophysiology data.
Such phenotypes depend on relationships among annotated events: onsets, offsets, durations, latencies,
ordering, co-occurrence, and persistence. QEL expresses these relationships using displacement,
bounded existence, and bounded universality, together with Boolean connectives and quantification
over rational variables.

The theoretical contribution is the LINE semantics and its related model checking algorithm. 
Finite interval
ensembles represent atomic annotations, and inverse translation, complement, future-directed
dilation, and erosion compute the supports of quantifier-free formulas. Closure and correctness
of these operations justify model evaluation directly on endpoint arrays. Quantifier elimination
extends the method to arbitrary formulas and finite evidence sets support representative truth assessment
of quantifier blocks. For a fixed formula, model checking has polynomial data complexity.
The separate semi-algebraic closure result establishes a corresponding spatial semantics over
Euclidean supports.

The GTC and base PGES examples show how one formula can state an ordered pattern together with
persistence requirements. They also distinguish the constraints a formula enforces from the
clinical interpretation supplied by annotations. The Cohort Builder makes formulas and matching
evidence accessible through a graphical timeline interface. In the illustrativel run, the PGES
query over 1,976 patients returned 37 patients and 45 annotation patterns in 15.97 seconds. On 54
manually constructed records, the base and stricter queries matched 43 and 33 records,
respectively, and both agreed with every assigned record-level outcome. These experiments
support feasibility and agreement on tested cases but establish neither general
implementation correctness nor clinical validity. In fact, the endpoint-only prototype is not a full-scale implementation,
and has the limitation that it is highly dependent on annotation quality.

In general, our formal results rest on well-formed LINE records. Annotation accuracy, completeness, and
boundary conventions are additional requirements before one can interpret logical matches with clinical confidence.
Further work includes query planning and indexing, comparisons with existing temporal-query
systems, clinical adjudication of selected cohorts, and semantics for uncertain annotations.
Spatial-temporal extensions and reusable libraries of validated phenotype specifications are
also natural directions.

Despite these limitations,  QEL provides a pathway toward  TRUER knowledge: 
a paradigm transforming approach  to make
biomedical knowledge an explicit, executable scientific object that humans and machines can inspect, challenge,
verify, and reuse, so that TRUER becomes a property of the biomedical data ecosystem by design
and by construction: transparent, because objects, relations, and evidence chains are explicit; rigorous, because
semantics are formal and reasoning is machine-verified; understandable, because a clinician can read
a formula; explainable, because every verdict carries its witnesses and its proof; and reproducible, because a
formula is re-executed, not re-implemented.

\section*{Code and data availability}

The synthetic evaluation set and executable checker (Section~\ref{sec:synthetic-evaluation}) are
available at \url{https://github.com/yhc29/QEL-demo} (commit \texttt{29e9028}). The clinical records belong to the Center for
SUDEP Research~\citep{csrpaper,csrweb} and are not included in this release.

\section*{Acknowledgment}\label{ack}

This work has been supported in part by the U.S. National Science Foundation Award IIS-2500624 and the National Institutes of Health grants R01AG084236 and U24AG098157. The views expressed in this paper are those of the authors and do not necessarily  reflect those of the funding agencies.
\printcredits

\section*{Ethics statement}
This study used only synthetic or previously acquired datasets and did not involve new recruitment, intervention, or
collection of identifiable human-subject data by the authors. Accordingly, no new institutional ethics approval or informed-consent procedure was required for this study. The work was conducted in accordance with the terms of use and licenses of the source datasets.

\section*{Declaration of competing interest}
The authors declare that they have no known competing financial interests or personal relationships that could have appeared to
influence the work reported in this paper.


\appendix
\section{Background and Proofs}\label{app:proofs}

This appendix opens with the algebraic background of
Section~\ref{subsec:qe-finite-evidence} (Appendix~\ref{app:doag}). It then gives the proofs that the main text defers, those of
Theorems~\ref{thm:sa-closure}, \ref{thm:finite-test-set}, and~\ref{complexity}. They combine the
results of Sections~\ref{sec:qel-rational} and~\ref{mc} with standard facts from real algebraic
geometry, the combinatorics of hyperplane arrangements, and the theory of linear arithmetic. The
appendix also establishes the two results on the limits of the problem that are cited in
Sections~\ref{subsec:prior-el} and~\ref{subsec:complexity}: the complexity of model checking when
the formula is part of the input (Proposition~\ref{prop:combined}) and the undecidability of
satisfiability (Proposition~\ref{prop:sat}).

\subsection{Rational Time as an Ordered Group}\label{app:doag}

The development of Section~\ref{subsec:qe-finite-evidence} rests on four properties of
rational time and on three classical facts about them. This subsection states both in elementary
terms.

\emph{Four properties.} QEL handles time points and durations by addition, subtraction, and
comparison, and the rational numbers support these operations as follows.
\begin{enumerate}
\item \emph{Shifts can be combined and undone.} The laws $(a+b)+c=a+(b+c)$, $a+0=a$, and
  $a+(-a)=0$ hold for all rational numbers. A set with an operation $+$, an element $0$, and an
  opposite $-a$ of each element $a$ is a \emph{group} if it obeys these laws. Among the
  non-negative numbers only $0$ has an opposite, and over non-negative time a displacement cannot
  refer to the past (Section~\ref{subsec:qel}).
\item \emph{The order of shifts is immaterial.} We have $a+b=b+a$; the group is \emph{abelian}.
\item \emph{Equal parts exist.} For every $a$ and every integer $n\ge1$ there is a $b$ with
  $nb=a$, where $nb$ is the sum of $n$ copies of $b$; the group is \emph{divisible}. Time in
  whole seconds, the group $\mathbb{Z}$, is not: no whole number of seconds is half of one second.
\item \emph{Shifts preserve earlier and later.} The relation $<$ is a linear order: it is
  transitive, and exactly one of $a<b$, $a=b$, and $b<a$ holds. Moreover, $a<b$ implies
  $a+c<b+c$; the group is \emph{ordered}. Time of day, which starts again at midnight, is a
  divisible abelian group that admits no such order.
\end{enumerate}
A group with these four properties and more than one element is a \emph{divisible ordered abelian
group}. It is \emph{dense}: for $a<b$, the midpoint $(a+b)/2$, which exists by divisibility, lies
strictly between $a$ and $b$. Hence the name \emph{dense ordered divisible abelian group}. The
rationals $(\Q,+,<)$ form the smallest such group that contains the integers, and the reals
$(\R,+,<)$ form another.

\emph{The first-order theory.} The language $\mathcal{L}_{\mathrm{oag}}$ of ordered abelian
groups (Lemma~\ref{lem:qe-doag}) has variables, which stand for elements of the group; the
symbols $0$, $+$, $-$, $<$, and $=$, and a constant for each rational number; the Boolean
connectives; and the quantifiers $\exists$ and $\forall$. The language is \emph{first-order}
because its quantifiers range over elements and not over sets of elements.
An integer multiple $nx$ abbreviates a sum, so that $2x<y+3$ is a formula of the language, whereas
the product $xy$ of two variables is not. A formula without free variables is a \emph{sentence},
and the \emph{first-order theory} of a structure is the set of sentences that are true in it.
Density, for example, is expressed by the sentence
\[
\forall x\,\forall y\,\bigl(x<y\rightarrow\exists z\,(x<z\wedge z<y)\bigr),
\]
which is true in $(\Q,+,<)$ and false in $(\mathbb{Z},+,<)$.

\emph{Three classical facts.} The first is \emph{quantifier elimination}
(Lemma~\ref{lem:qe-doag}): in a dense ordered divisible abelian group, every formula is
equivalent to one without quantifiers. The basic step is the elimination method of Fourier and
Motzkin for linear inequalities. Each comparison that mentions $z$ is solved for $z$, which
divisibility permits, and an element strictly above given lower bounds and strictly below given
upper bounds exists iff every lower bound lies below every upper bound. Thus
$\exists z\,(x<z\wedge z<y)$ is equivalent to $x<y$, by density; over $\mathbb{Z}$ the
equivalence fails for $x=3$ and $y=4$.
The second is \emph{completeness}: all dense ordered divisible abelian groups have the same
first-order theory~\citep{marker2002}. No sentence of the language distinguishes rational time
from real time, and nothing is lost by computing with rational numbers. The third concerns
\emph{decidability and complexity}.
The theory is decidable in exponential space~\citep{ferrante1975}, and every decision procedure
takes exponential time on infinitely many sentences~\citep{fischerrabin1974}. The cost is due to
the alternation of quantifiers: sentences with a bounded number of alternations lie in the
polynomial hierarchy~\citep{sontag1985}. Multiplication changes the picture. The first-order
theory of $(\Q,+,\cdot)$ is undecidable~\citep{robinson1949}, and the reduction in the proof of
Proposition~\ref{prop:combined} would carry this over to model checking if the terms of QEL
admitted products of variables. 

They are built from addition and the norm, which over $\Q$ is the absolute
value and is definable from the order (Section~\ref{syntax}).

Quantifier elimination yields Proposition~\ref{prop:qel-semilinear} and the finite evidence
sets of Theorem~\ref{thm:finite-test-set}; completeness and the complexity bounds yield
Proposition~\ref{prop:combined}. Theorem~\ref{complexity} shows that the cost is
polynomial in the size of the record once the formula is fixed.

\subsection{Proof of Theorem~\ref{thm:sa-closure}}\label{app:sa}

Let $A\subseteq\R^n$ be defined by an ordered-ring formula $\Phi_A(y)$ with real parameters.
By Tarski--Seidenberg quantifier elimination~\citep{tarski1951decision,seidenberg1954new,bochnak1998real,basu2006algorithms},
every set definable by such a formula is semi-algebraic. It suffices to give formulas for the
stated operations.

The complement is defined by $\neg\Phi_A(x)$, and displacement by $v\in\R^n$ by
$\Phi_A(x+v)$, where $x+v$ abbreviates $(x_1+v_1,\ldots,x_n+v_n)$. Dilation is defined by
\[
\exists y_1\cdots\exists y_n\,\Bigl(\Phi_A(y)\wedge\sum_{i=1}^n(x_i-y_i)^2\le r^2\Bigr),
\]
since $x\in A\oplus B_r$ iff some $y\in A$ is within distance $r$ of $x$. Erosion is defined by
\[
\forall z_1\cdots\forall z_n\,\Bigl(\sum_{i=1}^n z_i^2\le r^2\ \rightarrow\ \Phi_A(x+z)\Bigr),
\]
which asserts $x+B_r\subseteq A$. Alternatively, closure under erosion follows from complement,
dilation, and Lemma~\ref{lem:euclidean-duality}. Each formula is in the language of ordered rings.
Treating $r$ as a variable and conjoining $r\ge0$ defines the corresponding set of pairs $(x,r)$.
Replacing $\le r^2$ by $<r^2$ gives the open-ball versions, including the empty ball at $r=0$.
Thus all the stated sets are semi-algebraic. \qed

\subsection{Proof of Theorem~\ref{thm:finite-test-set}}\label{app:finite}

\emph{Evidence sets.} By Proposition~\ref{prop:qel-semilinear}, satisfaction at $q$ is determined
by the signs of finitely many affine polynomials. Include the polynomials defining the declared
domains. Each sign cell then lies wholly within or outside the admissible domain, and truth is
constant on it. Choose one representative of every admissible cell. Any satisfying assignment has
a satisfying representative, proving the existential equivalence. Any falsifying assignment has
a falsifying representative, proving the universal equivalence.

\emph{Computability and cardinality.} Lemmas~\ref{lem:qe-doag} and~\ref{lem:definability}
compute the affine polynomials. A nonempty cell is a rational solution set of linear equations and
strict inequalities, so it has a computable rational representative. In fixed dimension $n$, an
arrangement of $m$ proper hyperplanes has $O((m+1)^n)$ cells, including all faces. For $n\ge2$,
standard arrangement construction, augmented with rational interior representatives, takes
$O((m+1)^n)$ arithmetic operations~\citep[Chapters~1 and~7]{edelsbrunner1987}; for $n=1$, sorting
the boundary points takes $O((m+1)\log(m+2))$. If $m=0$, the sole cell is $\Q^n$.

\emph{Number of hyperplanes.} Fix the formula. Its translation has length $O(N)$: every atomic
occurrence contributes two constraints per interval, and the number of auxiliary and original
quantifiers depends only on the formula. The test-term elimination described in
Appendix~\ref{app:complexity} increases its length by a polynomial factor at each elimination.
After a fixed number of eliminations, the length, and hence the number of affine boundary
hyperplanes, remains polynomial in $N$.

\emph{Quantifier-free formulas.} Let $E$ contain the finite endpoints for the atoms of
$\varphi$. We construct $\mathcal H$ directly by tracking the terms that shift these endpoints.
For each subformula $\theta$, define its \emph{left} and \emph{right shift forms} by
\[
\begin{array}{@{}l@{}}
\Lambda_L(p)=\Lambda_R(p)=\{0\};\\[1pt]
\Lambda_L(\neg\theta)=\Lambda_R(\theta),\quad \Lambda_R(\neg\theta)=\Lambda_L(\theta);\\[1pt]
\Lambda_\bullet(\theta\wedge\theta')=\Lambda_\bullet(\theta\vee\theta')=\Lambda_\bullet(\theta)\cup\Lambda_\bullet(\theta');\\[1pt]
\Lambda_\bullet(\theta_u)=\{\ell+u\mid\ell\in\Lambda_\bullet(\theta)\};\\[1pt]
\Lambda_L(\mdiamond_t\theta)=\{\ell+t\mid\ell\in\Lambda_L(\theta)\},\quad
\Lambda_R(\mdiamond_t\theta)=\Lambda_R(\theta);\\[1pt]
\Lambda_L(\boldBox_t\theta)=\Lambda_L(\theta),\quad
\Lambda_R(\boldBox_t\theta)=\{\ell+t\mid\ell\in\Lambda_R(\theta)\},
\end{array}
\]
where $\bullet\in\{L,R\}$, and put $\Lambda(\theta)=\Lambda_L(\theta)\cup\Lambda_R(\theta)$.
Each form sums displacements and radii along a syntactic path. Each set has at most as many forms
as $\theta$ has atomic occurrences.

\begin{lemma}[Endpoints of supports]\label{lem:endpoint-forms}
Let $\theta$ be a subformula of a quantifier-free formula $\varphi$, with admissible values of the
variables. Every finite left endpoint of the canonical form of the support
$\semantics{\theta}_\varepsilon$ is of the form $e-\ell$ with $e\in E$ and
$\ell\in\Lambda_L(\theta)$, and every finite right endpoint is of the form $e-\ell$ with $e\in E$
and $\ell\in\Lambda_R(\theta)$.
\end{lemma}

\begin{proof}
Induct on $\theta$. Canonicalization selects existing endpoints; complementation exchanges left
and right endpoints. Intersection selects one of the two left endpoints and one of the two right
endpoints, while union introduces none. These observations prove the Boolean cases. Displacement
subtracts $u$ from every endpoint. By Proposition~\ref{prop:ie-closure}, dilation subtracts $t$
from left endpoints and erosion subtracts $t$ from right endpoints. These are exactly the
recurrences for $\Lambda_L$ and $\Lambda_R$.
\end{proof}

Let $\mathcal H_\varphi$ contain the proper hyperplanes
\begin{equation}\tag{A.1}\label{eq:family}
\begin{array}{@{}l@{\quad}l@{}}
\ell(\mathbf{x})=e-q & (e\in E,\ \ell\in\Lambda(\varphi)),\\[1pt]
\ell(\mathbf{x})-\ell'(\mathbf{x})=e-e' & (e,e'\in E,\ \ell,\ell'\in\Lambda(\theta)),\\[1pt]
x^i=0,\quad u(\mathbf{x})=0,
\end{array}
\end{equation}
where $\theta$ ranges over subformulas, $x^i$ over the duration and nonzero variables, and $\|u\|$
over norm subterms. Expand each shift form into its finitely many linear sign choices, replacing
each $\|u\|$ by $u$ or $-u$, and omit constant equations. The first line compares endpoints with
$q$, the second compares endpoints with each other, and the third fixes the declared-domain
conditions and norm signs. Zero norm arguments remain permitted whenever the complete radius is
positive. Each cell is wholly admissible or inadmissible; on it every norm has a fixed linear
expression. For fixed $\varphi$, there are $O((|E|+1)^2)=O(N^2)$ hyperplanes.

We show that truth at $q$ is constant on each admissible cell. Represent an endpoint by a
\emph{symbol} $(e,\ell)$ whose value is $e-\ell$. Two assignments in the same cell give the same
order to all symbols used at any subformula, and the same order between $q$ and the final symbols.
Compute canonical supports under both assignments. By Lemma~\ref{lem:endpoint-forms}, every finite
endpoint has such a symbol. Intersection, erosion, and normalization decide emptiness and whether
unions form intervals by endpoint comparisons and closure marks. The comparisons agree on the
cell, and the closure marks follow the same fixed rules. Induction therefore gives identical
symbolic interval lists. The final membership test at $q$ also agrees, proving the claim.

\emph{The linear case.} Suppose every modality is applied to a closed formula $\psi$. Its support
is independent of $\mathbf{x}$, with endpoints $e-\ell$ for closed shift forms $\ell$. For a
canonical member with endpoints $c,d$ and closure relations $\lambda,\rho$,
Proposition~\ref{prop:ie-closure} gives the tests
\[
\begin{array}{ll}
\mdiamond_t\psi: & c-t<w\ \land\ w\,\rho\,d,\\
\boldBox_t\psi: & c\,\lambda\,w\ \land\ w+t\le d.
\end{array}
\]
The modality holds iff a member passes the corresponding test; infinite-endpoint constraints are
omitted. Outside the outermost modalities, only Boolean connectives, displacement, and atoms
remain. Consequently, truth at $q$ is a Boolean combination of these tests at $w=q+u$, and of
atomic interval tests. Every boundary compares $q$ with some $e-\ell$. The first and third lines
of~\eqref{eq:family} therefore suffice, giving $m=O(N)$.

\emph{The quadratic bound is attained.} For $K\ge2$, let $p$ hold exactly at $i/K$ and $p'$
exactly at $j$, for $0\le i,j<K$. This record has $\Theta(K)$ thresholds and boundary entries.
At $q=0$, the formula $\mdiamond_1(p\wedge p'_x)$ holds exactly at the $K^2$ distinct values
$x=j-i/K$. Each is an isolated satisfying point, so every sign-cell decomposition preserving truth
must include it as a boundary. Thus $m\ge K^2=\Omega(N^2)$. \qed

\subsection{Proof of Theorem~\ref{complexity}}\label{app:complexity}

\emph{Part~1.} Lemma~\ref{lem:definability} proves the translation correct, including declared
domains and norm graph variables. Lemma~\ref{lem:qe-doag} preserves equivalence under quantifier
elimination, proving Algorithm~\ref{alg:model_checking} correct.

For fixed $\psi$, the translation has length $O(N)$ and a fixed number of quantifiers. Eliminate
these from the inside out by test terms~\citep{ferrante1975}. For a quantifier-free matrix $\theta$,
solve its atomic conditions involving $z$ for their affine boundary values $b_1,\ldots,b_s$.
If there are none, $\theta$ is independent of $z$. Otherwise the terms
\[
b_i,\qquad (b_i+b_j)/2,\qquad b_i-1,\qquad b_i+1
\]
sample every boundary, every interval between consecutive distinct boundaries, and both exterior
rays, for every assignment of the remaining variables. Thus $\exists z\,\theta$ is equivalent to
the disjunction of the corresponding substitutions; universality uses their conjunction. Clear
positive denominators after substitution. The domain guard is already part of the matrix. Each
elimination uses $O(s^2)$ copies, so a fixed number of eliminations leaves polynomial length in
$N$. The resulting closed Boolean formula is evaluated in linear time in that length.

\emph{Part~2.} Theorem~\ref{thm:finite-test-set} reduces the block to its representatives, and
Corollary~\ref{cor:qf-support} proves the support evaluation correct. A satisfying representative
supplies explicit values for the quantified variables.

For fixed $\psi$, the explicit arrangement has $m=O(N^2)$ hyperplanes, or $O(N)$ in the linear
case. Its representatives are constructed in $O((m+1)^k)$ arithmetic operations for $k\ge2$,
with $O((m+1)\log(m+2))$ required for $k=1$. Each substituted subformula has $O(N)$ canonical
intervals: displacement, dilation, and erosion cannot increase their number; complement adds at
most one; and union or intersection of ensembles with $i$ and $j$ members has at most $i+j$
members. All operations are linear passes on sorted endpoint arrays. After one initial sort and
canonicalization costing $O(N\log N)$, each representative therefore costs $O(N)$ to evaluate.
Since $k\ge1$, sorting and arrangement construction are absorbed by $O(N^{2k+1})$ in general and
$O(N^{k+1})$ in the linear case.

\emph{Databases.} Put $N_i=2+$ the number of boundary entries in record $i$, so $N_i\ge2$ even
for an empty record. If $N=\sum_iN_i$, then $\sum_iN_i^c\le N^c$ for every $c\ge1$. Summing the
record-local costs gives the stated database bounds. \qed

\subsection{Combined Complexity and Satisfiability}\label{app:limits}

Theorem~\ref{complexity} fixes the formula. We now allow it as input, obtaining the combined
complexity of model checking~\citep{vardi1982complexity}, and then consider satisfiability over LINEs.

\begin{proposition}[Combined complexity]\label{prop:combined}
The problem of deciding whether $(\varepsilon,q)\models\psi$, for a given LINE $\varepsilon$,
reference point $q$, and closed QEL formula $\psi$, is interreducible in polynomial time with the
decision problem of the first-order theory of $(\Q,+,{<})$. Consequently, it is PSPACE-hard; there
is a constant $c>0$ such that it is not decidable in nondeterministic time $2^{cn}$, where $n$ is
the length of the input; and it is decidable in exponential space. The lower bounds hold for a
fixed LINE with two atomic propositions.
\end{proposition}

\begin{proof}
\emph{From model checking to arithmetic.} The translation $\widehat{\Phi}(q)$ of
Algorithm~\ref{alg:model_checking} is computed in polynomial time and is true in $(\Q,+,{<})$,
with rational constants, iff $(\varepsilon,q)\models\psi$ (Lemma~\ref{lem:definability}).
Auxiliary variables define term values, including absolute values, without duplicating subformulas. Constants are removed as follows. Write them as integer multiples $k_1\kappa,\ldots,k_r\kappa$ of
one positive rational $\kappa$. For every rational $\lambda>0$, the map $x\mapsto\lambda x$ is an
automorphism of $(\Q,+,{<})$; the sentence is therefore equivalent to
$\exists w\,(w<w+w\wedge\sigma)$, where $\sigma$ has the multiple $k_iw$ in place of the $i$th
constant, and these multiples are defined in polynomial length by repeated doubling, with
auxiliary variables.

\emph{From arithmetic to model checking.} Let $\varepsilon_0$ be the LINE with the two temporal
events $(Z,\{[0,0]\})$ and $(B,\{(-\infty,0)\})$, and let $q=0$. For sums $u,v$ of variables, put
\[
\begin{array}{@{}l@{}}
(u<v)^\ast:=\exists w\,(Z_{v+w}\wedge B_{u+w}),\\[2pt]
(u=v)^\ast:=\exists w\,(Z_{v+w}\wedge Z_{u+w}),
\end{array}
\]
with one reserved variable $w$, locally bound in every translation. At $0$, $Z_{v+w}$ holds iff $w=-v$,
and then $B_{u+w}$ holds iff $u<v$, and $Z_{u+w}$ holds iff $u=v$. Replacing each atom
of a sentence $\sigma$ by its starred version gives a closed QEL formula
$\sigma^\ast$, of length $O(|\sigma|)$, such that $(\varepsilon_0,0)\models\sigma^\ast$ iff
$\sigma$ is true.

\emph{Bounds.} The theory of non-trivial divisible ordered abelian groups is
complete~\citep{marker2002}, so $(\Q,+,{<})$ and $(\R,+,{<})$ have the same theory. It is decidable
in space $2^{O(n)}$~\citep{ferrante1975}, and there is a constant $c>0$ such that no
nondeterministic algorithm decides it in time $2^{cn}$~\citep{fischerrabin1974}; the polynomial forward reduction therefore gives space $2^{\operatorname{poly}(n)}$, while the
linear-size reverse reduction preserves the exponential-time lower bound. PSPACE-hardness follows by translating a quantified Boolean
formula over $\varepsilon_0$:
a variable $x$ is restricted to the values $0$ and $1$ by relativizing its quantifier to
$Z_x\vee Z_{x+(-1)}$, the literal $x$ is rendered as $Z_{x+(-1)}$, and $\neg x$ as $Z_x$. The
resulting formula holds at $(\varepsilon_0,0)$ iff the quantified Boolean formula is true.
\end{proof}

\begin{proposition}[Satisfiability]\label{prop:sat}
No algorithm decides whether a closed QEL formula $\varphi$ is satisfied by some LINE, that is,
whether $(\varepsilon,q)\models\varphi$ for some LINE $\varepsilon$ and some $q\in\Q$. The set of
formulas with this property is recursively enumerable.
\end{proposition}

\begin{proof}
\emph{Enumeration.} Only the ensembles of the atomic propositions of $\varphi$ matter. Their
endpoints are rational or infinite, so the pairs $(\varepsilon,q)$ can be enumerated, and
Algorithm~\ref{alg:model_checking} decides for each pair whether it satisfies $\varphi$.

\emph{Undecidability.} By Trakhtenbrot's theorem, it is undecidable whether a first-order sentence
$\sigma$ in the vocabulary of one binary relation $G$, with equality, has a finite
model~\citep[Theorem~9.2]{libkin2004elements}. We translate $\sigma$ into a closed QEL formula
with two atomic propositions, $V$ and $R$. Let $\sigma^V$ be obtained from $\sigma$ by replacing
$G(x,y)$ by $R_{x+y+y}$; $x=y$ by $\forall w\,(V_{w+x}\leftrightarrow V_{w+y})$, with a fresh
variable $w$; $\exists x\,\theta$ by $\exists x\,(V_x\wedge\theta)$; and $\forall x\,\theta$ by
$\forall x\,(V_x\rightarrow\theta)$. Put
\[
\sigma^\ast:=\exists x\,V_x\ \wedge\ \forall x\,\forall d\,\neg(\boldBox_dV)_x\ \wedge\ \sigma^V,
\]
where $d$ is a duration variable.

Fix $\varepsilon$ and $q$, and let $A:=\{a\in\Q\mid q+a\in|\varepsilon(V)|\}$. The first conjunct
of $\sigma^\ast$ states that $A$ is non-empty. The second states that $|\varepsilon(V)|$ contains
no window $[r,r{+}d)$ with $d>0$; since $|\varepsilon(V)|$ is a finite union of intervals, this holds
iff $A$ is finite. Assume both. For $a,b\in A$, the formula
$\forall w\,(V_{w+a}\leftrightarrow V_{w+b})$ holds at $q$ iff the sets $A-a$ and $A-b$ coincide,
which implies $a=b$ by comparing their minima; the converse is immediate. By induction on the subformulas of $\sigma$, with
values in $A$ for the variables, $(\varepsilon,q)\models\sigma^V$ iff $\sigma$ holds in the finite
structure with universe $A$ and relation $\{(a,b)\in A^2\mid q+a+2b\in|\varepsilon(R)|\}$. Hence
$\sigma$ has a finite model whenever $\sigma^\ast$ is satisfied by a LINE.

Conversely, a finite model of $\sigma$ is isomorphic to a structure $(A,G)$ with universe
$A=\{4^i\mid 0\le i<n\}$. On $A\times A$ the map $(a,b)\mapsto a+2b$ is injective, as the digits
of $a+2b$ to the base $4$ show. The LINE with the temporal events $(V,\{[a,a]\mid a\in A\})$ and
$(R,\{[a{+}2b,a{+}2b]\mid(a,b)\in G\})$ therefore satisfies $\sigma^\ast$ at $q=0$.
\end{proof}

Validity over LINEs, that is, truth at every pair $(\varepsilon,q)$, is not recursively
enumerable. Otherwise the formulas $\varphi$ for which $\neg\varphi$ is valid, which are
the unsatisfiable ones, could be enumerated; run in parallel with the enumeration of
Proposition~\ref{prop:sat}, this would decide satisfiability. Thus no sound, complete,
effectively enumerable proof system exists for validity over LINEs.


\end{document}